%% file: main.tex
\documentclass[conference]{IEEEtran}
\usepackage{tcolorbox}

\usepackage{cite}
\usepackage{parcolumns}
\usepackage{amsmath,amssymb,amsfonts}
\usepackage{algorithmic}
\usepackage{graphicx}
\usepackage{textcomp}
\usepackage{xcolor}
\usepackage{lipsum}
\usepackage{framed}
\usepackage{soul}
\usepackage[export]{adjustbox}
\usepackage{floatrow}
\usepackage{hyperref}
\usepackage{algorithm}
\usepackage{subfiles}
\usepackage{verbatim}
\usepackage{listing}
\usepackage{listings}   
\usepackage{subcaption}
\usepackage{enumitem}
\usepackage{tikz}
\newcommand*\circled[1]{\tikz[baseline=(char.base)]{
\node[shape=circle,font=\bfseries,thick,draw=black,fill=black,text=white,inner sep=0.8pt] (char) {#1};}}

\usepackage[T1]{fontenc}
\usepackage{listings,xcolor,beramono}
\usepackage{tikz}
\usepackage{booktabs}
\makeatletter
\newenvironment{btHighlight}[1][]
{\begingroup\tikzset{bt@Highlight@par/.style={#1}}\begin{lrbox}{\@tempboxa}}
{\end{lrbox}\bt@HL@box[bt@Highlight@par]{\@tempboxa}\endgroup}

\newcommand\btHL[1][]{%
  \begin{btHighlight}[#1]\bgroup\aftergroup\bt@HL@endenv%
}
\def\bt@HL@endenv{%
  \end{btHighlight}%
  \egroup
}
\newcommand{\bt@HL@box}[2][]{%
  \tikz[#1]{%
    \pgfpathrectangle{\pgfpoint{1pt}{0pt}}{\pgfpoint{\wd #2}{\ht #2}}%
    \pgfusepath{use as bounding box}%
    \node[anchor=base west, fill=orange!30,outer sep=0pt,inner xsep=1pt, inner ysep=0pt, rounded corners=3pt, minimum height=\ht\strutbox+1pt,#1]{\raisebox{1pt}{\strut}\strut\usebox{#2}};
  }%
}

\newcommand{\code}[1]{\textsf{#1}}
\newcommand{\concept}[1]{\textbf{\textsf{#1}}}
\newcommand{\figref}[1]{Figure~\ref{#1}}
\newcommand{\secref}[1]{Section~\ref{#1}}
\newcommand{\listref}[1]{Listing~\ref{#1}}

\makeatother

\lstdefinelanguage{JavaExt}[]{Java}{
  frame=tlrb,
  basicstyle={\tiny\ttfamily},
  morekeywords={assert,constexpr,size_t},
  moredelim=**[is][{\btHL[fill=green!30,draw=red,dashed,thin]}]{@@}{@@},
  moredelim=**[is][{\btHL[fill=lightyellow!90,draw=white,dashed,thin]}]{|}{|},
  moredelim=**[is][{\btHL[fill=azure!90,draw=white,dashed,thin]}]{@|}{|@},
}

\def\BibTeX{{\rm B\kern-.05em{\sc i\kern-.025em b}\kern-.08em
    T\kern-.1667em\lower.7ex\hbox{E}\kern-.125emX}}
    
\definecolor{dkgreen}{rgb}{0,0.6,0}
\definecolor{gray}{rgb}{0.5,0.5,0.5}
\definecolor{mauve}{rgb}{0.58,0,0.82}
\definecolor{lightyellow}{rgb}{1,1,.6}
\definecolor{cyan}{rgb}{0.0, 1.0, 1.0}
\definecolor{azure}{rgb}{0.67, 0.9, 0.93}
\definecolor{carolinablue}{rgb}{0.6, 0.73, 0.89}
\hypersetup{
    colorlinks=true,
    linkcolor=blue,
    filecolor=blue,      
    urlcolor=blue,
}

\newcounter{mycomment}
\newcommand{\mycomment}[2][]{%
\refstepcounter{mycomment}%
{%
\setstretch{0.75}
\todo[inline,color={red!100!green!33},size=\scriptsize]{%
\textbf{[\uppercase{#1}\#\themycomment]:}~#2}%
}}
\newcommand{\task}[1]{\mycomment[Task]{#1}}

\renewcommand{\task}[1]{}

\usepackage{tikz}
\usepackage{textcomp}
\usepackage{hyperref}
\usepackage{lipsum}

\newcommand\copyrighttext{%
  \footnotesize \textcopyright 2021 IEEE. Personal use of this material is permitted. Permission from IEEE must be obtained for all other uses, in any current or future media, including reprinting/republishing this material for advertising or promotional purposes, creating new collective works, for resale or redistribution to servers or lists, or reuse of any copyrighted component of this work in other works.}
\newcommand\copyrightnotice{%
\begin{tikzpicture}[remember picture,overlay]
\node[anchor=south,yshift=10pt] at (current page.south) {\fbox{\parbox{\dimexpr\textwidth-\fboxsep-\fboxrule\relax}{\copyrighttext}}};
\end{tikzpicture}%
}

\begin{document}

\title{ArCode: Facilitating the Use of Application Frameworks to Implement Tactics and Patterns}

\author{
\IEEEauthorblockN{Ali Shokri}
\IEEEauthorblockA{
\textit{Rochester Institute of Technology}\\
Rochester, NY, United States \\
as8308@rit.edu}
\and
\IEEEauthorblockN{Joanna C. S. Santos}
\IEEEauthorblockA{
\textit{Rochester Institute of Technology}\\
Rochester, NY, United States \\
jds5109@rit.edu}
\and
\IEEEauthorblockN{Mehdi Mirakhorli}
\IEEEauthorblockA{
\textit{Rochester Institute of Technology}\\
Rochester, NY, United States \\
mxmvse@rit.edu}
}

\maketitle
\copyrightnotice{}
\IEEEpeerreviewmaketitle


\begin{abstract}
Software designers and developers are increasingly relying on application frameworks as first-class design concepts. They instantiate the services that frameworks provide to implement various architectural tactics and patterns. One of the challenges in using frameworks for such tasks is the difficulty of learning and correctly using frameworks’ APIs.  
This paper introduces a learning-based approach called \textsc{ArCode} to help novice programmers correctly use frameworks' APIs to implement architectural tactics and patterns. \textsc{ArCode} has several novel components: a graph-based approach for learning specification of a framework from a limited number of training software, a program analysis algorithm to eliminate erroneous training data, and a recommender module to help programmers use APIs correctly and identify API misuses in their program.  
We evaluated our technique across two popular frameworks: JAAS security framework used for authentication and authorization tactic and Java  RMI framework  used to enable remote method invocation between client and server and other object oriented patterns. 
Our evaluation results show (i) the feasibility of using \textsc{ArCode} to learn the specification of a framework; (ii) \textsc{ArCode} generates accurate recommendations for finding the next API call to implement an architectural tactic/pattern based on the context of the programmer's code; (iii) it accurately detects API misuses in the code that implements a tactic/pattern and provides fix recommendations. 
Comparison of \textsc{ArCode} with two prior techniques (MAPO and GrouMiner) on API recommendation and misuse detection shows that \textsc{ArCode} outperforms these approaches.
\end{abstract}

\begin{IEEEkeywords}
Software Framework, Architectural Tactics, API Specification, API Usage Model, API Recommendation, API Misuse Detection
\end{IEEEkeywords}

\input{Introduction}
\input{Overview}
\input{Preparation}
\input{Graam}

\input{SpecMiner}

\input{Recommendation}
\input{ExperimentalStudy}
\input{ThreatsToValidity.tex}
\input{RelatedWork}
\input{Conclusion}

\section*{Acknowledgments}
This work was partially funded by the US National Science Foundation (NSF) under grant number CCF-1943300, CNS-1816845 and CNS-1823246. 

\bibliographystyle{IEEEtran}
\bibliography{IEEEabrv,bibliography}

\end{document}

%% file: Introduction.tex
\section{Introduction}

To satisfy performance, security, reliability and other quality concerns, architects need to compare and carefully choose a combination of architectural patterns, styles or tactics. In the subsequent development phase, these architectural choices must be implemented completely and correctly in order to avoid a drift from envisioned design. 
Prior work~\cite{KazmanFrameworks} by Cervantes, Velasco-Elizondo, and Kazman confirms that software designers and developers are increasingly relying on application frameworks as first-class design concepts to facilitate implementation of architectural tactics and patterns.
\textit{Software frameworks} are reusable software elements that provide key functionalities, addressing recurring concerns across a range of applications. They incorporate many architectural patterns and tactics to prevent software designers and developers from implementing software from scratch~\cite{KazmanFrameworks,cervantes2019data}. For instance, the architecture of most contemporary enterprise applications relies on the \textit{Spring Framework} that provides pre-packaged solutions to implement various architectural concepts ranging from \textit{Model-View-Controller} (MVC) patterns to \textit{authentication} and \textit{authorization} security tactics~\cite{Bass,ICSA2017}.

Developers use Application Programming Interfaces (APIs) to import and use the frameworks' functionalities~\cite{cervantes2019data,MUJHID201781}. Therefore,  programs' quality largely depends on using these APIs correctly~\cite{ICSE2018}. Multiple studies have shown that proper use of a framework's API requires an in-depth understanding of its underlying architectural patterns and tactics, class structure, and set of tacit sequence calls, data-flows as well as interfaces that need to be implemented \cite{KazmanFrameworks,robillard2009makes,Johnson:1992:DFU:141936.141943,ICSE2018}. Recent qualitative and quantitative studies have 
reported that implementing architectural tactics is more complex compared to delivering software functionalities, and novice  and non-architecture savvy developers struggle in implementing architectural tactics and patterns~\cite{ICSE2018, ICSA2017}.
Previous paper by  Soliman, Galster, and  Riebisch published at ICSA~\cite{7930203} has indicated that developers rely on sources such as Q\&A websites (e.g. StackOverflow) to find information on how to use frameworks, implement tactics and patterns for specific quality attributes~\cite{DBLP:conf/wicsa/SolimanGSR16}. However, prior research  have also shed a light that snippets on accepted answers of Q\&A websites can contain design flaws, bugs or vulnerabilities that get reproduced across multiple software systems that reused that code snippet ``as is''~\cite{baltes2019usage,fischer2017stack}.

The prior work on framework API recommendations~\cite{acharya2007mining, shoham2008static, mover2018mining, fowkes2016parameter, dang2015api, lamba2015pravaaha, amann2019investigating,zhong2018empirical,zhong2009mapo,hsu2011macs,dang2015api,saied2018towards,saied2020towards,gu2019codekernel, thung2016api, zheng2011cross,ren2020demystify,monperrus2013detecting,amann2018systematic,amann2019investigating,wasylkowski2007detecting} focus on low level, local data structure related concerns and basic utility frameworks used to implement various data structures. 
This line of work fall short of addressing the challenges of implementing tactics and patterns and has not fully studied frameworks used to bring a new architectural tactic or pattern into a given system design. 

Other researchers have attempted to develop recommender systems to assist programmers in implementing architectural tactics and patterns~\cite{ICSA2017,DBLP:conf/icse/MirakhorliCK15,DBLP:conf/icsa/BhatTSBHM19}. However, prior work does not support implementing architectural tactics and patterns using frameworks.
In this paper, we aim to study frameworks with architectural implications that address tactics or patterns. 
We propose an approach entitled \textsc{ArCode} to help programmers implement an architectural tactic or pattern using API recommendations. \textsc{ArCode} leverages a novel learning technique to infer an \textit{accurate} and \textit{explicit} API specification model which will be used for generating recommendations. 


The significance of the contribution made by this paper is briefly described below:
\begin{itemize}[leftmargin=*]
\item To the best of our knowledge, this is the first study focusing on inferring and using API specification of a software framework with architectural implications (e.g., frameworks implements architectural tactics and patterns). 

\item We present a program analysis approach to reverse engineer a novel \textsc{\textbf{G}raph-based F\textbf{r}amework \textbf{A}PI Us\textbf{a}ge \textbf{M}odel} (\textbf{GRAAM}), which is an \textit{abstract and semi-formal representation} of how a framework's API is being used in a given program to implement tactics and patterns.

\item An automated approach to detect projects that violate frameworks API properties. We release a program analysis method that analyze the byte code of frameworks and extract an explicit usage model of framework APIs using a concept named \textsc{\textbf{I}nter-\textbf{F}ramework \textbf{D}ependency} (\textbf{IFD}) model. Later the IFD  can be used to identify projects that violate a framework's implicit API order constraints.
    \item A novel \textit{inference algorithm} called \textbf{\textsc{ArCode}} to construct the framework API specification model from a repository of \textbf{GRAAMs}. \textsc{ArCode} is an inter-procedural context-, and flow-  sensitive static analysis approach to automatically infer a specification model of frameworks from two sources, limited sample programs and a framework source code.
    
   \item An empirical investigation of the usefulness of the proposed approach to recommend APIs and detect API misuses for tactics' implementation on two popular Java-based frameworks.
  We use Java Authentication and Authorization Services (JAAS) and Remote  Method Invocation (RMI) as case studies. JAAS framework is used to implement \textit{authentication} and \textit{authorization} security tactics~\cite{ICSA2017}, while RMI framework is used as a building block to implement a number of object oriented patterns such as client-server and remote method call.
   We demonstrate that \textsc{ArCode} can help developers implement tactics and patterns via an accurate API recommendation and API misuse detection when such frameworks are used.

\end{itemize}

\begin{figure*}
    \centering
        \includegraphics[width=\textwidth]{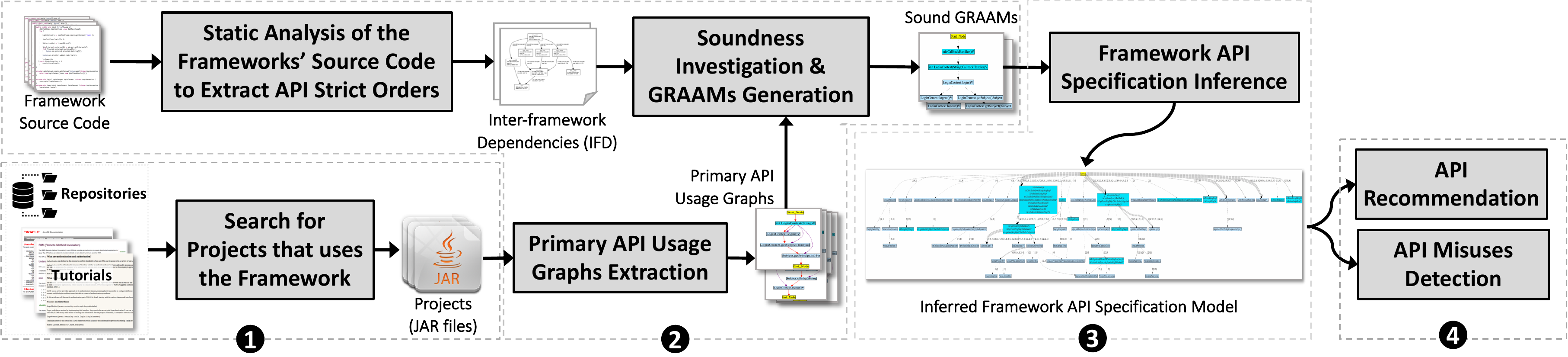}
        \centering
        \caption{Overview of our approach (\textsc{ArCode})}
        \label{fig:overview}
\end{figure*}

The remainder of this paper is organized as follows.  
\secref{sec:overview} provides an overview of our approach. \secref{sec:Preparation} briefly describes data collection phase.
\secref{sec:GRAAM} presents our graph-based API usage model (GRAAM) for a program. 
\secref{sec:ArCode} describes \textsc{ArCode}, our automated approach for inferring a framework API specification model from extracted GRAAMs. 
An experimental evaluation of the approach is presented in \secref{sec:experimentalStudy}. 
\secref{sec:Threats_to_Validity} discusses threats to validity of the approach.
\secref{sec:relatedWork} reviews related work.
Lastly, \secref{sec:conclusion} concludes this paper.

%% file: Overview.tex
\section{Overview}
\label{sec:overview}

\textsc{ArCode} is designed to perform API recommendation and misuse detection for application frameworks used to implement architectural tactics and patterns. In particular, these frameworks express a great degree of inter-process communication and API interactions beyond a single class, module or process~\cite{ICSA2017}.
\textsc{ArCode} aims to help novice developers, and non-architecture savvy developers to use frameworks. 
\textsc{ArCode} learns the API specification of frameworks from limited \textit{sound} program examples. A program is considered \concept{sound} if it uses the framework API correctly to implement the tactic.
As shown in \figref{fig:overview}, the approach has four phases:

\begin{itemize}[leftmargin=18pt]
    \item[\circled{1}]  \textbf{Data Collection and Preparation Phase:} We  create a code repository of programs that incorporate the framework of interest. Section~\ref{sec:experimentalStudy} use \textit{learning saturation} as a measure of indicating how many training projects are     required.

\item[\circled{2}] \textbf{Training-Data Pre-Processing and Validation Phase:}
Given the training data, we perform a \textit{context-}, \textit{path-}, and \textit{flow-sensitive} static analysis to create an \textit{inter-procedural} graph-based representation of API usages for each program. In this phase, we ensure that only sound training programs will be included in our training data.
To guarantee the syntax correctness, we first compile programs and generate their jar files. 
Moreover, to verify that a program is semantically error free (w.r.t. API usages), we use a novel technique to extract \textsc{\textbf{I}nter-\textbf{F}ramework \textbf{D}ependency} (\textbf{IFD}) model from the framework's internal source code. This model encompasses mandatory order constraints of APIs that are enforced by the framework and must be preserved in any given program. This approach identifies and rules out incorrect API usages and API violations. 
Leveraging the IFD model, we separate sound programs from those that violate implicit rules of API usage required by the framework.

\item[\circled{3}] \textbf{Training Phase:} 
For each sound program, we create its \textbf{G}raph-based F\textbf{r}amework \textbf{A}PI Us\textbf{a}ge \textbf{M}odel (GRAAM) which is an \textit{abstract model} demonstrating how the framework API was used in a given validated training data. Finally, via a recursive learning method the framework API specification model is created from a set of generated GRAAMs. 

\item[\circled{4}] \textbf{Recommendation Phase}:
The learnt model is used in this phase to implement a recommender system guiding programmers in using APIs. Specifically, this system analyzes the program under development and recommends what APIs should be considered and how they should be used in that program to correctly implement a tactic or a pattern. Furthermore, if there are API misuses, ARCODE will detect and recommend fixes.
\end{itemize}

%% file: Preparation.tex
\section{Data Collection and Preparation Phase}\label{sec:Preparation}

The first phase of this approach is focused on collecting a set of projects that uses a framework of interest. 
These projects should match the following criteria:
\textit{(i)} it imports and uses the framework API, and
\textit{(ii)} it is syntactically correct.
For checking the first condition, we scan the source code and check for API statements in the code.
To guarantee the syntax correctness (second condition), we compile the programs to generate their JAR files.

%% file: Graam.tex
\section{Training Data Processing and Validation Phase}\label{sec:GRAAM}

In the previous phase, we obtained collected projects that use a framework API and are compilable.
However, these two characteristics do not guarantee these projects to be  \textit{sound}.
Therefore, this second phase focuses on finding sound projects that can be used for training.
%
%

Distinguishing sound and unsound projects require a sophisticated \textit{\textbf{program representation model}} that can capture fundamental information regarding the usage of the framework APIs including but not limited to object instantiation, static and non-static method calls, as well as static and non-static field accesses.
These constructs can be scattered over multiple methods, which requires an inter-procedural analysis of the program.
Such a representation must also capture implementation of interfaces and inheritances of abstract classes of the framework. 
Moreover, this model shall be able to represent relationships between the framework APIs, allowable data dependencies between these methods, the logical and a usecase-driven order of method calls, and semantically equivalent sequences of API calls. 

In this phase, we present an approach to pre-process the training data to (1) extract the above elements, and (2) validate the training data so projects which incorrectly implement tactics/patterns can be eliminated from the training process.



\subsection{Motivating Examples}\label{subsec:Examples}
\listref{code:JAASSample1} and \listref{code:JAASSample2} show two real-world code snippets of the \textit{authentication} tactic~\cite{Bass} implemented using the JAAS framework. The correct implementation of this security tactic requires a careful sequence of API calls and manipulations of data objects. 
Although these two code snippets implement authentication tactic in a different code structure, they both: create a \code{Subject} object and an object that implements the \code{CallbackHandler} interface and pass them to the constructor of the instantiated
\code{LoginContext} object; and then call the \code{login()} method from \code{LoginContext}.

%
In addition to the above statements, the code in \listref{code:JAASSample2} goes further and calls the \code{getSubject()} method from a \code{LoginContext} object (line 11) and calls the \code{getPrincipal()} method from a \code{Subject} object (line 12).

From these two examples, we observe the followings:

\noindent- \textbf{Observation \#1:} Only a subset of statements in a program (the highlighted lines) are related to the framework of interest and should be part of a framework API specification model. 

\noindent- \textbf{Observation \#2:} The framework-related statements could be scattered across multiple methods. Thus, the framework API specification model shall be able to interconnect these statements that are within different scopes. 

\noindent- \textbf{Observation \#3:} Multiple (sub) programs with the same behavior might be written slightly different. 
For instance, both code snippets in \listref{code:JAASSample1} and \listref{code:JAASSample2} create \code{Subject} and \code{CallbackHandler} objects to be passed to the constructor of \code{LoginContext}. Although the order of object instantiation in two code snippets is different, it does not change the behavior of the program.

\begin{listing} 
\tiny 
\begin{lstlisting}[language = javaEXT]
1. public class TestJaasAuthentication {
2.    public static void main(String[] args) {
3.        String user = System.getProperty("user");
4.        String pass = System.getProperty("pass");
5.        boolean loginStatus = true;
6.        try {
8.            LoginContext loginContext = getLoginContext(user,pass);
9.           | loginContext.login(); | 
10.        } catch (LoginException e) { loginStatus = false; }
11.        if(loginStatus) System.out.println("Login Successful.");
12.        else System.out.println("Login Failed.");
13.    }
14.    private static LoginContext getLoginContext(String u, String p) throws LoginException{
15.       | CallbackHandler handler = new RanchCallbackHandler(u, p); |
16.       | Subject subject = new Subject(); |
17.       | LoginContext lc = new LoginContext("RanchLogin", subject, handler)}; | 
18.        return lc;
19.    }
20. }
\end{lstlisting}
\caption{\textit{Sample \#1: a code snippet that implements Authentication tactic using JAAS framework}}\label{code:JAASSample1}
\end{listing}
\normalsize

\begin{listing} 
\tiny
\begin{lstlisting}[language = JavaExt]
1. public class LoginUsecase {
2.    private static Logger LOGGER = Logger.getLogger(LoginUsecase.class);
3.    public static void main(String[] args){
4.        BasicConfigurator.configure();
5.        LoginContext lc = null;
6.        System.setProperty("java.security.auth.login.config", "jaas2.config");
7.        try{
8.            |Subject subject = new Subject();| 
9.            |lc = new LoginContext("rainyDay2", subject, new JAASCallbackHandler(| 
                |"user1", "pass1"));| 
10.           |lc.login();|
11.           |Subject subject = lc.getSubject();|
12.           |subject.getPrincipals();|
13.           LOGGER.info("established new logincontext");
14.        }
15.        catch (LoginException e){
16.            LOGGER.error("Authentication failed " + e);
17.        }
18.    }
19. }
\end{lstlisting}
\caption{\textit{Sample \#2: a code snippet that implements Authentication tactic using JAAS framework}}  \label{code:JAASSample2}
\end{listing}
\normalsize

Therefore, it is necessary to develop an API usage representation model that adequately captures correct usages of frameworks' APIs while taking these concerns into account. We developed a novel \textbf{Graph-based Framework API Usage Model (GRAAM)} to address these concerns.

\subsection{Pre-Processing and Validation}~\label{subsec:PrimaryAPIUsageGraph}
This pre-processing step identifies APIs that are used to implement tactics and patterns in a program, validates their usage based on several ground-truths obtained from the framework's source code, and creates \textit{Graph-based Framework API Usage Models (GRAAMs)} for correct API usages.

We follow a four-step process:
\textbf{(1)} System Dependence Graph (SDG) extraction,
\textbf{(2)} Slicing of the SDG,
\textbf{(3)} Removal of API usage violations, and
\textbf{(4)} Generation of Graph-based Framework API Usage Models (GRAAMs).

\subsubsection{System Dependence Graph (SDG) Extraction}
First, we perform an \textit{inter-procedural} static analysis on a program and extract its context-sensitive \textit{Call Graph} using 1-CFA algorithm. A call graph is a directed graph representing relationships between caller and callee methods in a program~\cite{grove1997call}. A 1-CFA context-sensitive callgraph~\cite{shivers1991control} distinguishes between situations where a method $m_3$ is being called from $m_1$ or $m_2$.
This type of call graph considers the possibility of different behaviors of the program in callee method (e.g.  \textit{$m_3$}) based on different caller methods (e.g. $m_1$ or $m_2$).
For the sake of scalability, we did not choose a higher sensitivity of context (i.e., n-CFA for $n>1$).



Next, we compute the System Dependence Graph (SDG) of the program under analysis. An SDG is a directed graph representing a whole program~\cite{horwitz1990interprocedural}. The nodes in this graph are \textit{statements} in the program and the edges are either \textit{data} or \textit{control} dependencies between nodes. 


Since we incorporate a context-sensitive call graph, the constructed SDG holds the following characteristics: 
\begin{itemize}[leftmargin=*]
    \item[--] \textit{flow-sensitive}: it accounts the order of execution of statements in the program being analyzed;
    \item[--] \textit{context-sensitive}: it distinguishes different call sites. The same method \textit{m} can be invoked by different methods (call sites). As a result, \textit{m} is analyzed differently based on its corresponding call site.
    \item[--] \textit{inter-procedural}: it represents the system as a whole, interconnecting statements within different methods based on the caller-callee relationships;
\end{itemize}

We use T. J. Watson Libraries for Analysis (WALA)\footnote{http://wala.sourceforge.net} to construct SDGs with the aforementioned attributes. We chose WALA over other tools (e.g. Soot~\cite{vallee2010soot}) because it provides built-in supports for extracting SDGs from a 1-CFA call graph as well as different versions of Java language (e.g. Java 8).

\subsubsection{Slicing the Extracted SDG}
In this second step, we compute a slice\footnote{A program slice includes only the set of statements that may affect a point of interest of the program (referred as the slice criterion)~\cite{weiser1981program}.} of the program \textit{p} under analysis that includes all statements \textit{s} in the SDG that are either (i) a  \textit{framework-related} statement; or (ii) that may be \textit{affected by} a framework-related statement  or (iii) that \textit{affects} a framework-related statement.
%
%
\concept{Framework-related Statements} are the statements $s_f$ that match the conditions (a)-(f) listed below. Each statement $s_f$ can be related with the framework either \textit{\textbf{directly}} (cases \textit{a}, \textit{b}, and \textit{c}) or \textit{\textbf{indirectly}} via inheritance (cases \textit{d}, \textit{e}, and \textit{f}): 
\begin{enumerate}[leftmargin=15pt]
    \item[(a)] it invokes a method from a \textit{framework data type} (classes or interfaces declared in the framework);
    \item[(b)] it instantiates an object from a \textit{framework data type};
    \item[(c)] it accesses a field from a \textit{framework data type};
    \item[(d)] it invokes a method implemented by an application class that inherits or implements a \textit{framework data type};
    \item[(e)] it instantiates an object from an application class that inherits or implements a \textit{framework data type};
    \item[(f)] it accesses a field from an application class that inherits or implements a \textit{framework data type};
\end{enumerate}

After computing a slice of the SDG, we remove all the nodes (statements) in the remained graph that are not framework-related yet keeping the direct and indirect dependencies between framework statements. 
The outcome of this process is the program's \concept{Primary API Usage Graph} $g = (V,E)$, 
%
%
which is a directed labelled sub-graph of the SDG, with nodes $v \in V$ and edges $e \in E$ where $E \subseteq V \times V$. 
The set of nodes $V$ in a primary API usage graph is partitioned into three types: start node $V_{start}$, end nodes $V_{end}$ and framework API usage nodes $V_f$:
\begin{enumerate}[leftmargin=15pt]
    \item[(1)] A \textit{\textbf{start node}} $v_{start} \in V_{start}$ represents the begining of the framework usage in a program. Each primary API usage graph starts with a single \textit{start node};
    \item[(2)] Each \textit{\textbf{end node}} $v_{end} \in V_{end}$ indicates the termination of the framework usage. Each primary API usage graph ends with one or more  \textit{end node(s)}; 
    \item[(3)] Each \textit{\textbf{framework API usage node}} $v_f \in V_f$ denotes a framework-related statement in a program (i.e., $V_f = S_f$). Each $v_f$ has an associated \textbf{\textit{instruction type}} $type(v_f)$ (i.e. \textit{object instantiation}, \textit{field access} or \textit{method invocation}). Furthermore, each $v_f$ has a \textbf{\textit{target framework data type}} $target(v_f)$, that depends on the instruction type. In case of object instantiations, $target(v_f)$ is the framework class type of the object; for method invocations, $target(v_f)$ is the target of the call; and for field accesses, the $target(v_f)$ is the framework type that contains the field. 
\end{enumerate}

The edge set $E$ in a primary API usage graph has two partitions: \textit{\textbf{sequence}} edges $E_s \subseteq (V_{start} \times V_f) \cup (V_f \times V_{end}) \cup (V_f \times V_f) $ and \textit{\textbf{data dependency}} edges $E_d \subseteq V_f \times V_f$:

\begin{enumerate}[leftmargin=15pt]
    \item[(1)] A data dependency edge $e_{data} = v_{src} \xrightarrow{data}  v_{dst}$ indicates that $v_{dst}$ uses data that has been defined by $v_{src}$. 
    \item[(2)] A sequence dependency $e_{seq} = v_{src} \xrightarrow{seq.}  v_{dst}$ indicates that $v_{dst}$ is used after $v_{src}$ in the program. 
\end{enumerate}

\begin{figure}
    \centering
    \begin{subfigure}[b]{0.4\textwidth}
        \centering
        \hspace{-1pt}
        \vspace{20pt}
        \includegraphics[width=.8\textwidth]{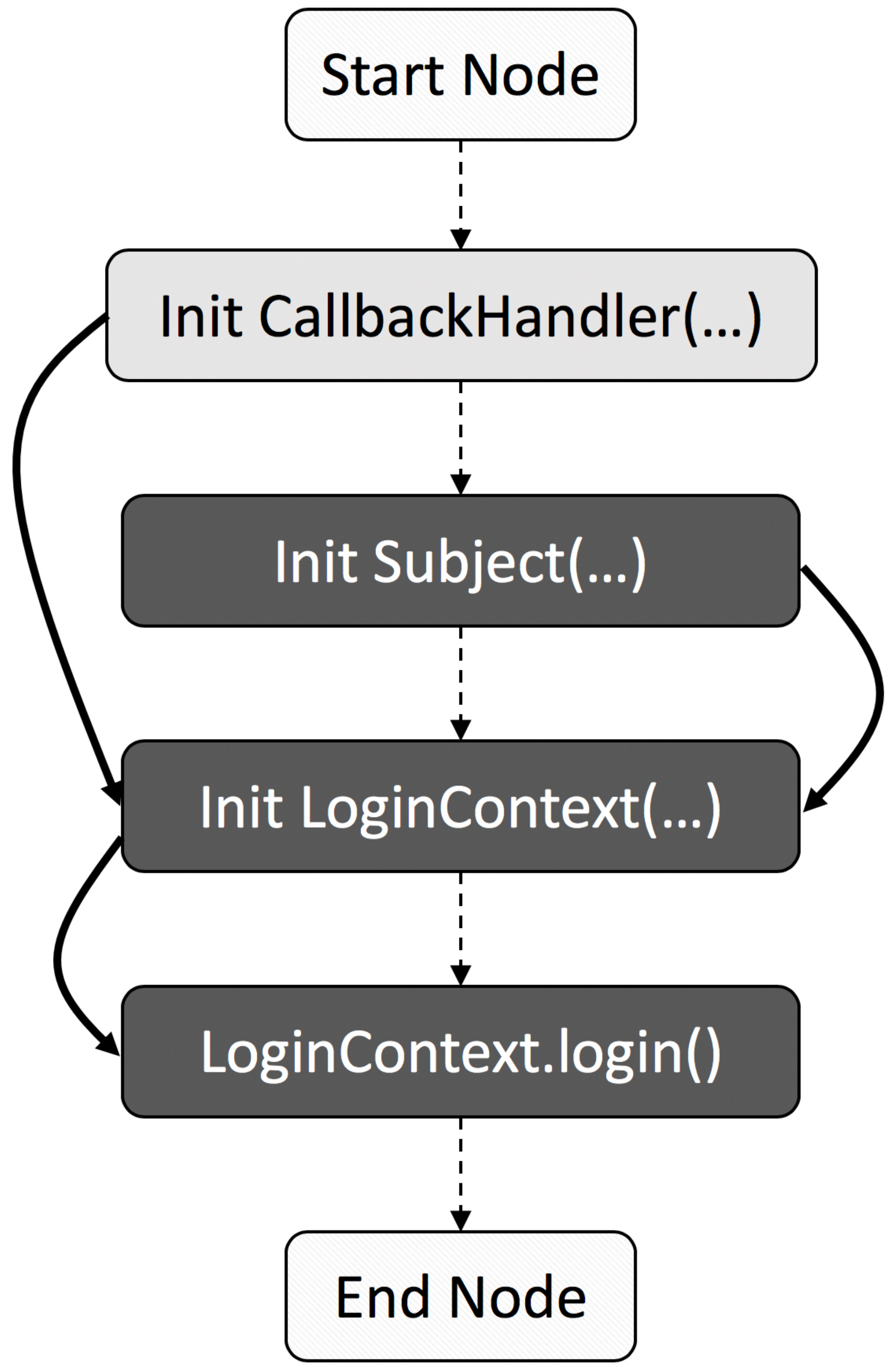}
        \centering
        \caption{Primary API usage graph for   \listref{code:JAASSample1}}
        \label{fig:Grasp_for_Sample1}
    \end{subfigure}%
    ~
    \begin{subfigure}[b]{0.4\textwidth}
        \centering
        \includegraphics[width=.8\textwidth, right]{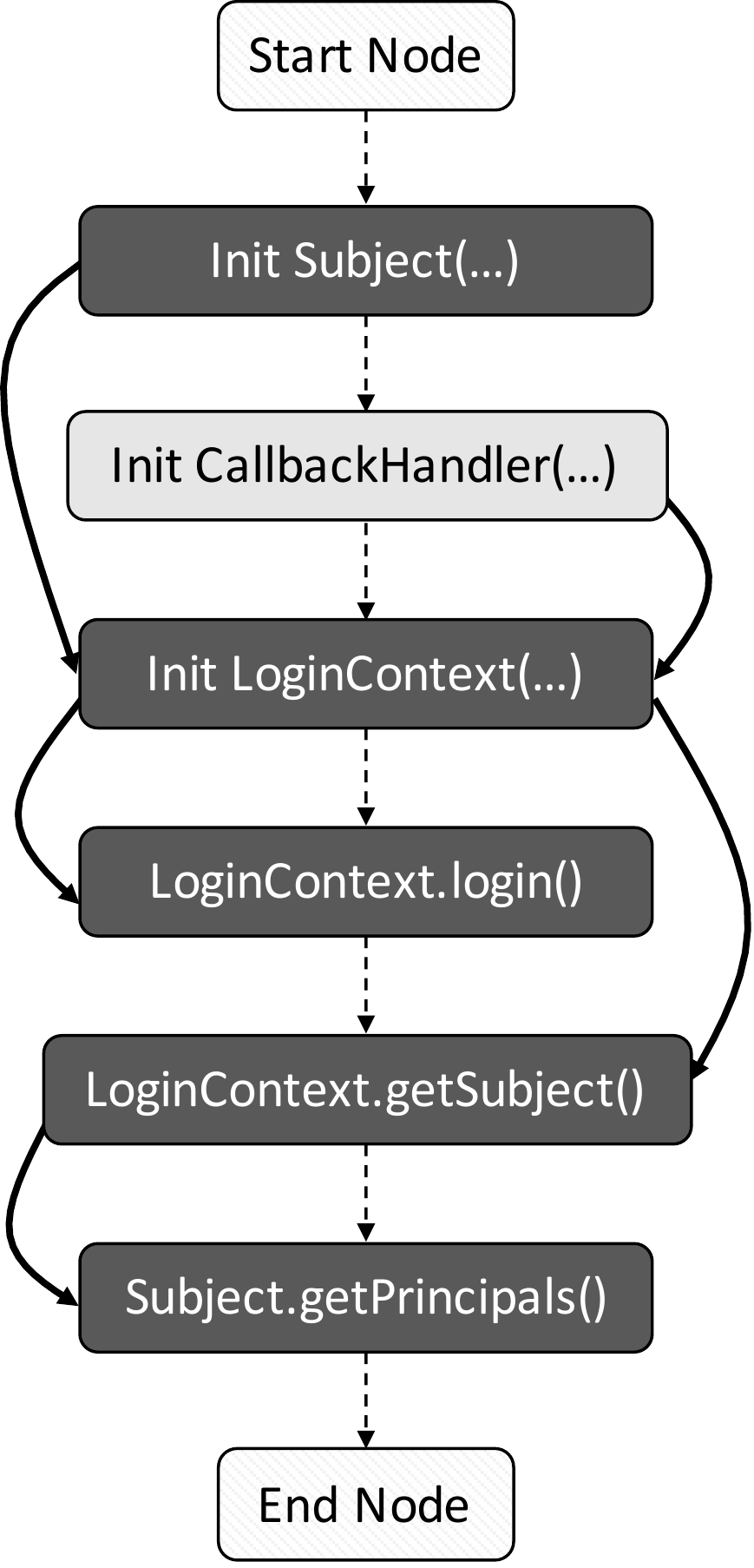}
        \caption{Primary API usage graph for \listref{code:JAASSample2} }
        \label{fig:Grasp_for_Sample2}
    \end{subfigure}
    \caption{Extracted primary API usage graphs from sample code snippets}
    \label{fig:Grasps_for_Sample_Codes}
        \vspace{-6pt}

\end{figure}

\figref{fig:Grasp_for_Sample1} and \figref{fig:Grasp_for_Sample2} show the primary API usage graphs for the code snippets in \listref{code:JAASSample1} and \ref{code:JAASSample2}.  
We use solid edges to show data dependencies and dashed edges to indicate sequence dependencies.  For example, since the \code{CallbackHandler} object  is used in the \code{LoginContext}'s constructor, there is a data dependency edge between \code{init CallbackHandler} and \code{init LoginContext} nodes in the primary API usage graphs. 
Nodes with a lighter background represent statements that are related with the framework indirectly (via inheritance). 

\subsubsection{Removing API Usage Violations}
We explained so far how we analyze a project to extract its API usage information and represent it as a primary API usage graph. However, we want to eliminate any erroneous API usages in our training corpus. Therefore, we need a reliable ground truth to identify such incorrect usages in a code repository and filter them out from the training data.


\textsc{ArCode} analyzes the framework's source codes and captures the API usage rules that are not visible to the developers, but are implicitly reflected in the source code of frameworks. To do so, it performs a static analysis of the framework's source code to capture  \textit{implicit} data dependencies between APIs of a framework. Since this information is obtained directly from the framework, not from programs that use the framework, it could be considered as a ground truth for identifying API misuses. 

The main idea is to leverage \textbf{reader-writer} roles of API methods inside a framework to find dependencies between them. Using these dependencies, one can find partial API strict orders that must be followed in a program. As a results, API misuses in a program could be identified with confidence by finding violations from these strict orders. \textit{Writer} and \textit{Reader} methods are defined as:

\noindent-- \textbf{\textit{Writer method:}} A method is considered as a writer regarding a specific class \textit{field} if it changes the value of that specific field somewhere in its body.

\noindent-- \textbf{\textit{Reader method:}} A method is considered as a reader regarding a specific class \textit{field} if it uses the value of that specific field somewhere in its body.

\begin{figure}[!htbp]
    \centering
    \includegraphics[width=\textwidth]{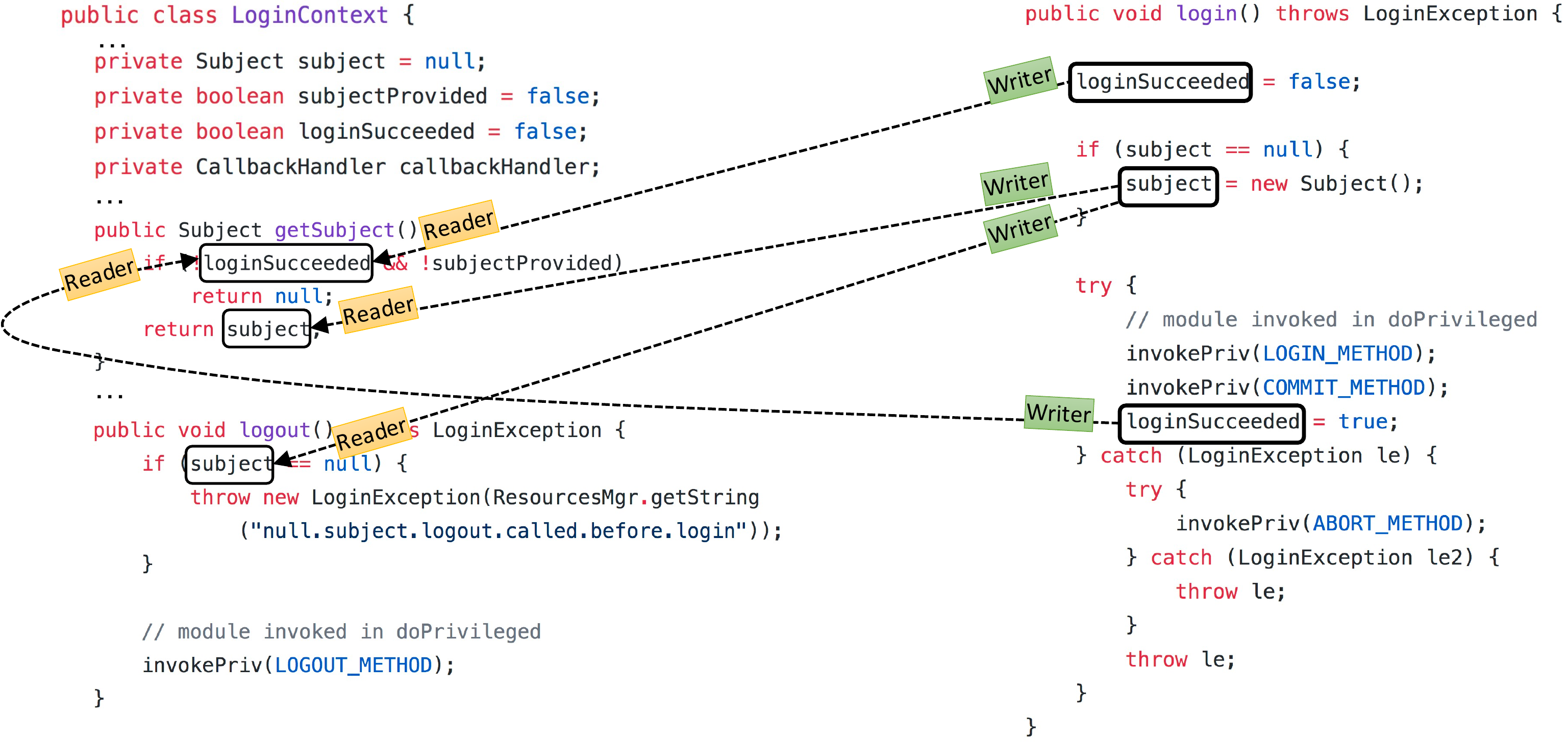}
    \caption{Extracting Inter-framework Dependencies between getSubject(), logout(), and login() methods which are three API methods inside LoginContext class (JAAS framework)}
    \label{fig:IFD_Sample}
        \vspace{-6pt}
\end{figure}

\begin{figure}[!htbp]
    \centering
    \includegraphics[width=.7\textwidth]{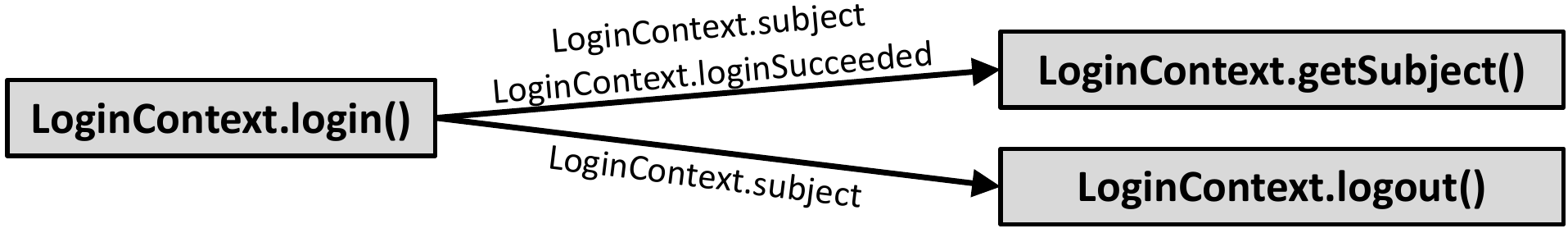}
    \caption{IFD Model based on
    Fig.~\ref{fig:IFD_Sample}}
    \label{fig:IFD_model}
        \vspace{-6pt}
\end{figure}

As an example, \figref{fig:IFD_Sample} shows three methods from \code{LoginContext} class in the JAAS framework. This class has various fields including \code{subject} and \code{loginSucceeded}. While method \code{login()} assigns values (\textit{writes}) to \code{subject} and \code{loginSucceeded} fields, methods \code{getSubject()} and \code{logout()} use (\textit{read}) these fields' values. As a result, \code{getSubject()} and \code{logout()} methods are dependent to some data generated in \code{login()} method. We extract these information and create our \concept{Inter-Framework Dependency (IFD)} model. 
The IFD built for the framework's code shown in \figref{fig:IFD_Sample} is depicted in \figref{fig:IFD_model}.




We use IFD model for identifying incorrect programs w.r.t. a framework usage. 
If a programmer has a framework-related statement
$s_2$ in a program before API $s_1$ while based on IFD model $s_2$ reads data generated by $s_1$ ($s_1 \xrightarrow[]{\text{data}} s_2 $), then, that program is considered as an incorrect program. 


\subsubsection{Generating Graph-based Framework API Usage Models (GRAAM)}
The last step of pre-processing and data validation phase focuses on creating a representation of each training data, a Graph-based Framework API Usage Model (GRAAM), which can be used by \textsc{ArCode}'s learning algorithm. We provide a definition for a GRAAM further in this section. 


As discussed earlier, it is possible that two programs with different sequences of framework-related statements implement the same tactic. For example, although \listref{code:JAASSample1} and \listref{code:JAASSample2} implement the same use case, the instantiation order of \code{Subject} and \code{CallbackHandler} in their programs and so, in their corresponding primary API usage graphs (\figref{fig:Grasp_for_Sample1} and \figref{fig:Grasp_for_Sample2}) is different. 
We call these sequences of API usage nodes as \concept{Semantically Equivalent API Sequences}. 
%
%
%
%
Two sequences of framework usage nodes, $seq_1$ and $seq_2$, are semantically equivalent if both conditions are true:
\begin{itemize}[leftmargin=*]
    \item 
    there is a \textit{bijection} between the nodes in $seq_1$ and $seq_2$.
    Each paired nodes $s_1 \in seq_1$ and $s_2 \in seq_2$ have the same \textit{type} and \textit{target framework type}, i.e. $type(s_1) = type(s_2)$ and $target(s_1) = target(s_2)$; and
    \item the two sequences are isomorphic considering only the data dependency between APIs in each sequence.
\end{itemize}

As a result,  a \concept{Graph-based Framework API Usage Model} (GRAAM) is a directed graph $g$ which has the same set of nodes as a \textit{primary API usage graph}, but a different edge type: \textbf{API Order Constraint} edges. Semantically equivalent API sequences have a single representative in a GRAAM. In other words, two different programs with the same behavior (w.r.t. APIs of the same framework) have isomorphic GRAAMs.

For example, parts of the programs in \listref{code:JAASSample1} and \listref{code:JAASSample2} including instantiation of \code{Subject}, \code{CallbackHandler}, \code{LoginContext}, and calling \code{login()} method, implement the same tactic (authenticate actors). Thus, their corresponding sub-GRAAMs are the same. This property of GRAAM enables us to have the same representation for semantically equivalent API usages in different programs. Hence, when we look at a code repository, we can correctly identify the same API usages and compute their frequencies.

\begin{figure}
    \centering
        \includegraphics[width=\textwidth]{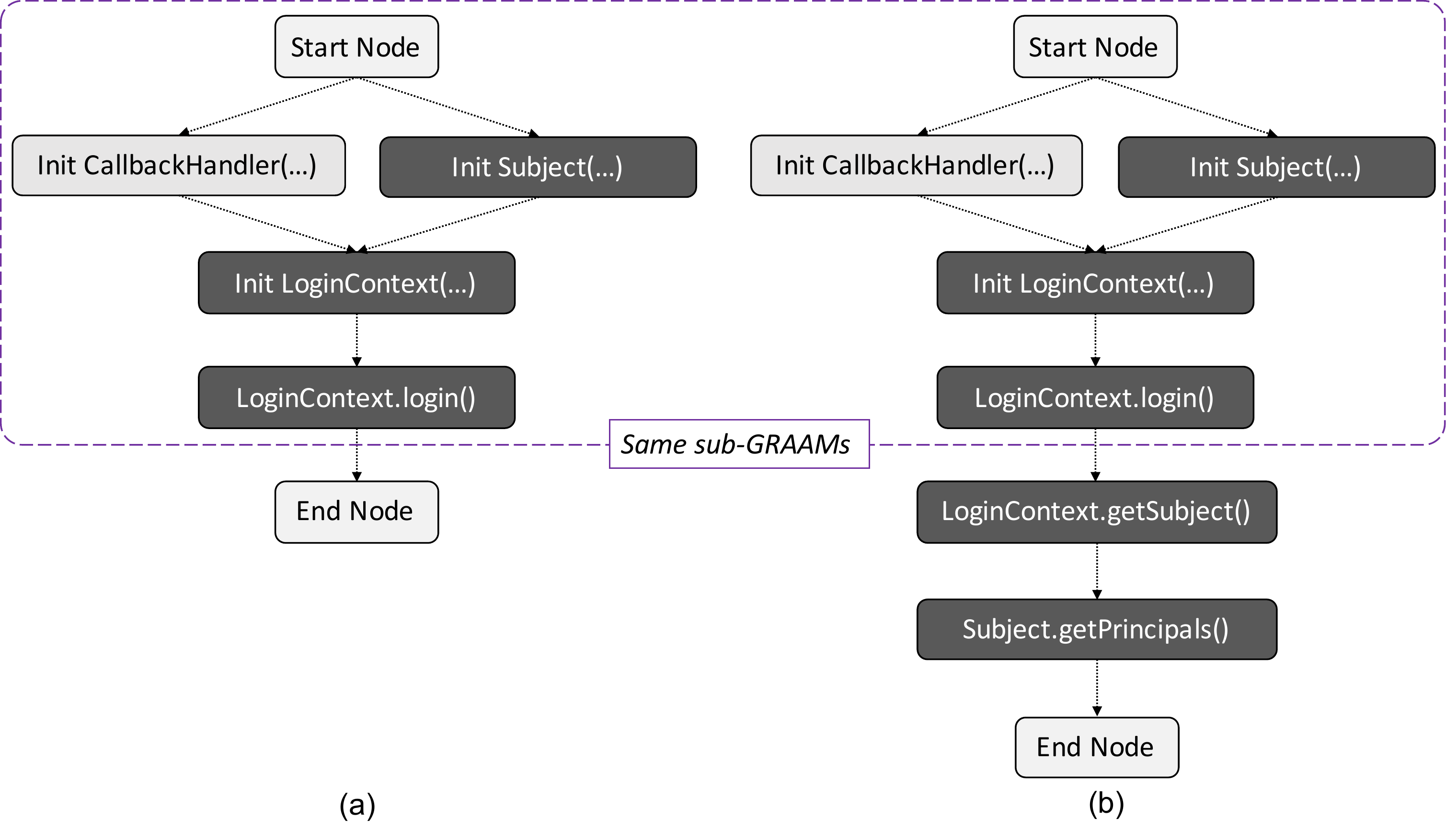}
        \caption{Built GRAAMs for programs in (a) \listref{code:JAASSample1} and (b) \listref{code:JAASSample2}. Since both programs use \textit{semantically the same} API sequences for creating a \code{LoginContext} object and calling \code{login()} method, their corresponding Sub-GRAAMs confined by the purple dashed line are isomorphically the same.}
        \label{fig:CreatedGRAAMs}

\end{figure}

To create a GRAAM, we \textbf{(i)} start with a validated primary API usage graph, \textbf{(ii)} remove all the sequence edges except edges from start node and edges to end node(s), and \textbf{(iii)} add relevant edges from IFD model to the graph. To be more specific, this graph is built based on framework-related statements identified in the program, their \textit{data dependencies} captured from the program, plus data dependencies mined from the framework's source code. 
The reasoning behind removing sequence edges and keeping data edges is that if the data produced in an API call $a_1$ is not being used by an API call $a_2$, it means that calling $a_1$ before $a_2$ in the program is not required, i.e., the order of $a_1$ and $a_2$ does not affect the program's behavior. We reflect this non-restriction situation in our GRAAM by eliminating this order constraint between $a_1$ and $a_2$.

\figref{fig:CreatedGRAAMs} shows the created GRAAMs for the examples in \listref{code:JAASSample1} and \ref{code:JAASSample2}. 
These GRAAMs are created based on the primary API usage graphs depicted in \figref{fig:Grasps_for_Sample_Codes} and the IFD model shown in \figref{fig:IFD_model}. For instance, the edge between \code{LoginContext.login()} and \code{LoginContext.getSubject()} in \figref{fig:CreatedGRAAMs}b comes from data dependency between \code{LoginContext.login()} and \code{LoginContext.getSubject()} captured in IFD model depicted in \figref{fig:IFD_model}. As shown in \figref{fig:CreatedGRAAMs}, parts of programs in \listref{code:JAASSample1} and \listref{code:JAASSample2} that implement the same thing have the same sub-GRAAMs.

%% file: SpecMiner.tex
\section{Training Phase: Inferring  a  Framework API Specification  Model  (FSpec)} 
\label{sec:ArCode}

\textsc{ArCode} uses the repository of created GRAAMs to infer the Framework API Specification Model (FSpec). 

%


\subsection{Framework API Specification Model (FSpec)}
\label{subsec:FSM}
We use the collected sound GRAAMs to build a unified graph-based Framework API Specification Model (FSpec) which represents  \textbf{correct} ways to use  the framework. 
This model aims to (a) reflect only possible correct combinations of API calls, (b) to contain only paths that represent a correct framework's API call sequence, and (c) to create one representative for all the semantically equivalent API usages.

 
 \begin{figure}
    \centering
\includegraphics[width=0.63\textwidth]{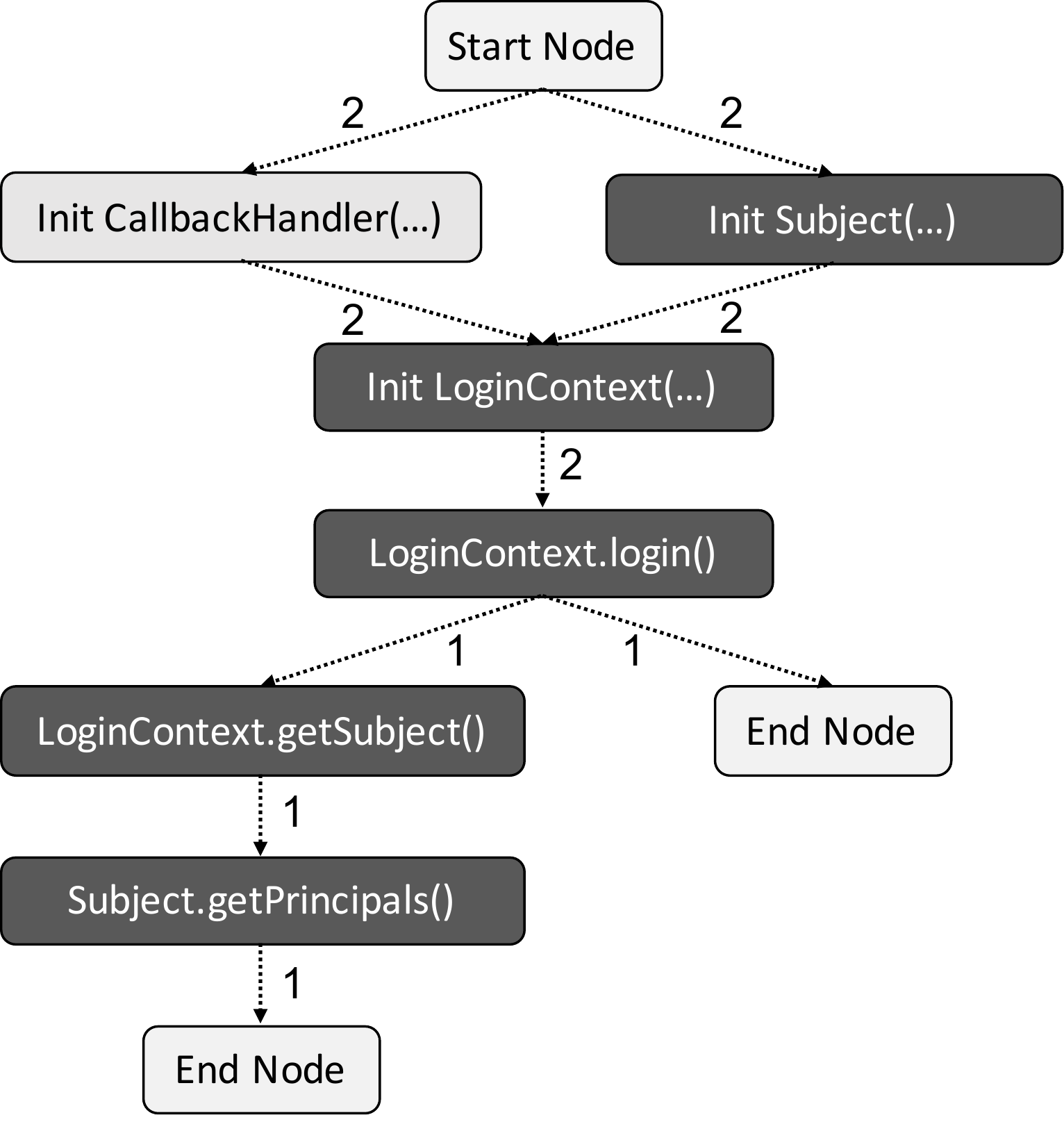}
  \caption{Built Framework API Specification Model (FSpec) from depicted GRAAMs in \figref{fig:CreatedGRAAMs}a and \figref{fig:CreatedGRAAMs}b}
  \label{fig:BuiltModel_1}
  \vspace{-6pt}
\end{figure}

 \figref{fig:BuiltModel_1} shows a Framework API Specification model (FSpec) built from the GRAAMs shown in \figref{fig:CreatedGRAAMs}a and \figref{fig:CreatedGRAAMs}b.
 Similar to a GRAAM, an FSpec encompasses three types of nodes: \textit{start node}, \textit{end node}, and \textit{framework-related node}. An FSpec has the same edge type as a GRAAM, \textit{API order constraint} edge. This type of edge represents the strict orders of framework APIs one should follow to correctly incorporate that framework in a program. However, FSpec has a new label on each edge: \textit{frequency}. Frequency of a sub-graph of FSpec represents the number of times that the corresponding API usage was observed in the code repository of sound programs.

For instance, all edges between the start node and \code{login()} node in  \figref{fig:BuiltModel_1} are labeled with frequency of $2$. That means, there were two (semantically) equivalent API usages observed in the code repository that both \textbf{(i)} instantiated objects of \code{Subject} and \code{CallbackHandler}, \textbf{(ii)} used those objects to instantiate an object of \code{LoginContext}, and then, \textbf{(iii)} called \code{login()} method of \code{LoginContext}. After the \code{login()} node, however, the frequencies are changed to 1. It means that there was a program that has not used any other APIs of JAAS framework after \code{login()} method. In addition to that, the model shows that there was a program that continued calling framework APIs and had two more API calls. 

FSpec represents \textit{semantically equivalent} API usages in a single representation.
This property shows the \textbf{generalizability} of \textsc{ArCode} in the sense that when it visits an instance of correct API usage in a program, a representative of all the semantically equivalent API usages to the originally visited one would be added to the FSpec under construction.

\subsection{Inferring a FSpec Model from Sound GRAAMs}
To infer a framework's API specification model (FSpec), \textsc{ArCode} finds mergeable parts of created \textit{sound} GRAAMs. We will define mergeable parts of GRAAMs later in this section. Through the inference process, if \textsc{ArCode} finds meargable part of a GRAAM similar to one that was previously added to the FSpec model, then, it  increases the frequencies of the corresponding edges in FSpec. Otherwise, it adds the corresponding new nodes to the model and sets their edge frequency to 1. 


To guarantee that all paths from start to end nodes represents a correct framework usage, we have provided inference rules to identify \concept{Mergeable sub-GRAAMs}.  Two sub-GRAAMs are mergeable if \textbf{(i)} both include the start node, and \textbf{(ii)} their corresponding sequences are \textit{semantically equivalent}.  
Assuming that there are two GRAAMs  $g_{1}$ and $g_{2}$, we explain how the merging algorithm works such that $g_{2}$ merges into $g_{1}$.

\begin{itemize}[leftmargin=*]
    \item We first identify all the mergeable sub-GRAAM pairs from $g_{1}$ and $g_{2}$;
    \item Amongst the identified merging candidates, the pair with the highest number of nodes is selected;
    \item The frequency of edges from $sub-g1$ increments by the frequency of corresponding edges from $sub-g2$;
    \item The remained parts of $g2$ which are not included in $sub-g2$ will be added to $g_1$;
    \item We repeat this process until there are no more mergeable sub-GRAAM pairs from $g_{1}$ and $g_{2}$. This process guarantees that in the end, no semantically equivalent sequences exists in the start node's children list.
\end{itemize}

\begin{figure}
    \centering
\floatbox[{\capbeside\thisfloatsetup{capbesideposition={left,center},capbesidewidth=4cm}}]{figure}[\FBwidth]
{\caption{Emergence of unexpected and unsound sequences by merging $g_{1}$ and $g_{2}$ from their middle parts}\label{fig:middle_merge}}
{\includegraphics[height=38mm]{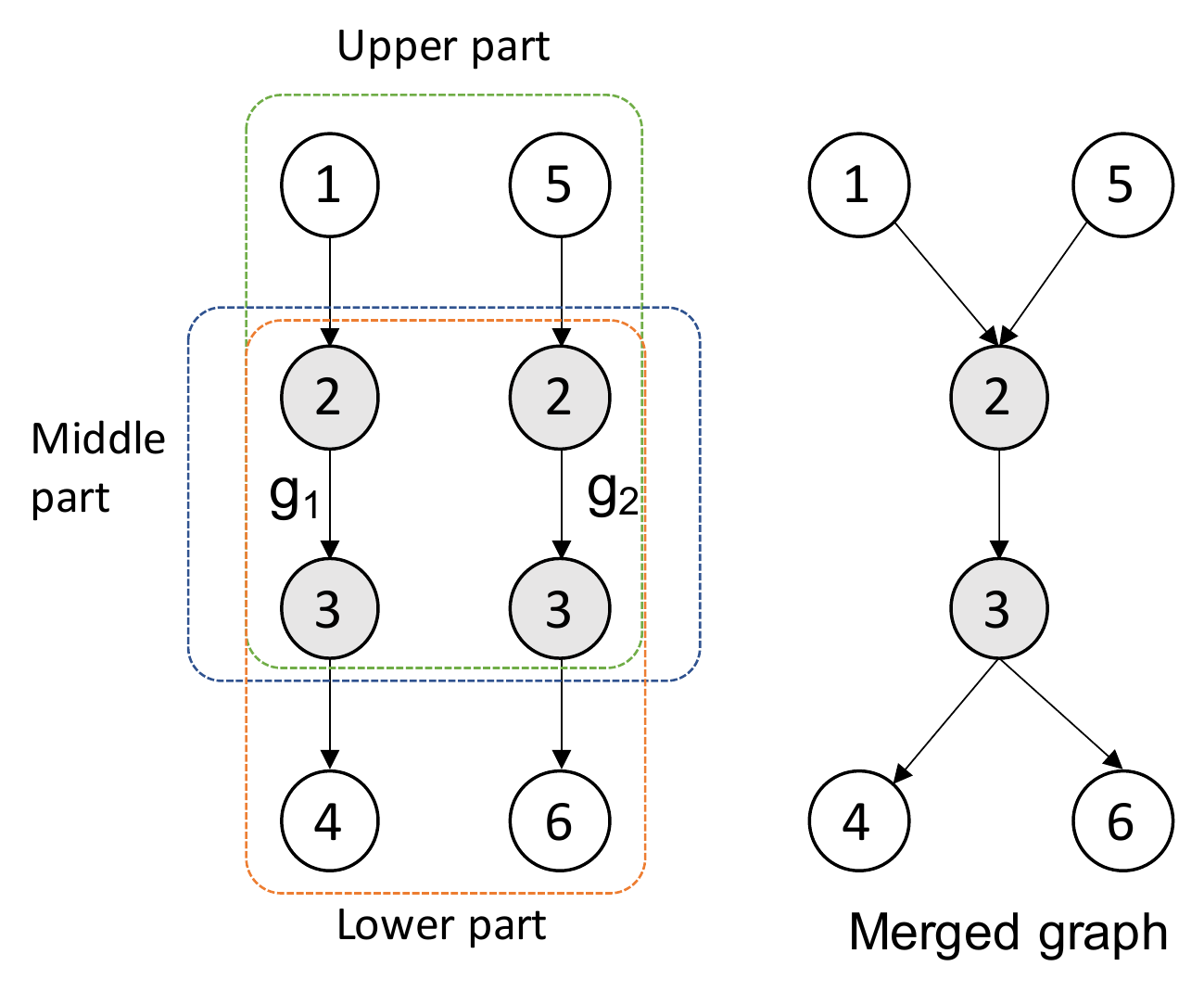}}
\end{figure}

To clarify this algorithm, there are three different parts in \figref{fig:middle_merge} that could be a candidate for merging purposes. The \textit{upper part} starts with the root (e.g. node $1$ from $g_1$ and node $5$ from $g_2$) of the graph and may (or may not) contain some successors of the root. The \textit{middle part} does not contain root nor any end nodes. The \textit{lower part} contains at least one end node and may (or may not) include predecessors of the end node.
To merge two GRAAMs $g_1$ and $g_2$, the algorithm only considers their \textit{upper parts} to avoid the emergence of incorrect paths. 
\figref{fig:middle_merge} provides an example of this situation. Assume that $1 \rightarrow 2 \rightarrow 3 \rightarrow 4$ and $5 \rightarrow 2 \rightarrow 3 \rightarrow 6$ are two \textit{correct} API sequences. Upon merging $g_1$ and $g_2$ from their middle part, two \textit{incorrect} paths ($1 \rightarrow 2 \rightarrow 3 \rightarrow 6$ and $5 \rightarrow 2 \rightarrow 3 \rightarrow 4$) are appeared in the merged graph. 


%% file: Recommendation.tex
\section{Recommendation Phase}
Once \textsc{ArCode} is trained and a FSpec model is built, it can be used to help programmers correctly implement architectural patterns and tactics through providing correct API recommendations. 
The recommendation system has the following steps:
\begin{enumerate}[leftmargin=*]
    \item \textbf{Process Partial Program:} it takes in a partial program written by a programmer and  creates its GRAAM;
    \item \textbf{Context Based Recommendation:} the recommender engine finds the most similar \textit{semantically equivalent} API usages inside the FSpect to the given GRAAM. Then it finds changes needed to be performed on the GRAAM to make it a correct implementation of a tactic or pattern.  
  \item \textbf{Ranked List}: the outcome is provided in the form of ranked list of API recommendations (e.g. remove, add, replace). The rank of each recommendation in the list is determined based on the frequency of its corresponding edges in the FSpec.
\end{enumerate}

\textsc{ArCode} can identify the next APIs required to be called in a program to make it a complete and correct implementation. It also is able to detect misuses of APIs in a program and recommend fixes for it. Some API misuses are not detectable in the compile time since it does not violate syntax of the language. However, these are serious semantic bugs which can compromise the entire objective of a tactic/pattern. 

%% file: ExperimentalStudy.tex
\section{Experimental Study}\label{sec:experimentalStudy}

To show the practical usefulness of our approach for complex software systems, we used it in experimental studies of JAAS (Java Authentication and Authorization Services) and RMI (Remote Method Invocation) frameworks to generate recommendations for projects that incorporate them to implement tactic and patterns.
JAAS addresses security architectural concerns in applications, and RMI supports interactions between different modules and methods in distributed systems.
In these two studied frameworks  \textbf{(i)} we investigate whether \textsc{ArCode} can learn correct ways framework APIs are invoked to implement tactics/patterns. We rely on a learning saturation experiment;  \textbf{(ii)} we evaluate the accuracy of \textit{API recommendation} to help programmers implement tactics and patterns, and \textbf{(iii)} we analyze ArCode's performance in \textit{API misuse detection}.  

Additionally, we compare \textsc{ArCode}'s performance with  MAPO~\cite{zhong2009mapo} and GrouMiner~\cite{Nguyen2009}, two approaches which were developed for API recommendation and misuse detection, but not evaluated on frameworks with architectural implications~\footnote{Frameworks used to implement tactics or patterns}.

\subsection{Data Collection}\label{subsection:DataCollection}
To create the training data for \textsc{ArCode}, we identified a large number of popular open-source projects using JAAS and RMI frameworks. These projects were collected from different public and open-source code repositories including GitHub, BitBucket and Maven. A team of three members peer reviewed programs and compiled them to generate their bytecode. The purpose of reviewing programs was to make sure that each selected program implemented authentication and authorization tactics and inter-process communication patterns and  had more than one API call in its code. 
The final repository contains \textbf{51} projects that uses JAAS, and \textbf{50} projects that uses RMI.
This repository has a total \textbf{1,106,886} lines of code (Java files), \textbf{8,831} classes and \textbf{9,600} methods.  

\subsection{Finding API Usages in Programs}\label{sec:evaluation_GRAAMs}
We created GRAAMs for projects in the code repository based on the approach discussed in this paper (\secref{sec:GRAAM}). We also created API usages for GrouMiner, and MAPO approaches accordingly. It is worth mentioning that a program may have more than one entrypoint~\footnote{A program's entrypoint is the first method invoked once it starts executing (e.g. the \textsf{main()} method).}. Therefore, it is possible that more than one API usage can be found per program. 
%

One issue we encountered in our experiments was failure of GrouMiner when the number of nodes in its API usage graph (i.e. graph-based object usage model) precedes 19. Since authors of GrouMiner provided their source code, and to stay faithful to the code, we did not make any changes to its source code. Therefore, we exclude API usage graphs with more than 19 nodes from our experiment.

\subsection{\textsc{ArCode} Learning Saturation}
\label{subsection:FSMCreation}

To investigate the learning capability of \textsc{ArCode}, we create its \textbf{\textit{learning curve}}. This curve represents the rate of new knowledge added to FSpec comparing to the size of new GRAAMs it visits. In the case that \textsc{ArCode} has a good learning ability, the rate of growth of FSpec should decrease by visiting more GRAAMs.

\figref{fig:model_saturation_JAAS} and \figref{fig:model_saturation_RMI} show learning curve of \textsc{ArCode} while creating FSpec for JAAS and RMI frameworks in the training phase. Blue curves represent the cumulative size of $k$ visited GRAAMs by ArCode. Orange curves show the size of FSpec after visiting $k$ GRAAMs.
To conduct this experiment, we first sort the GRAAMs based on the number of their nodes. Then, we feed \textsc{ArCode} starting with bigger chunks of information.
While there is an increase in the number of FSpec nodes in the initial steps, after visiting 24 GRAAMs of JAAS-based programs, we observe reaching  90\% completion of the final FSpec model. It means that after this point, there would be only 10\% new information learnt by FSpec. Likewise, \textsc{ArCode} reaches 90\% completion of its FSpec after visiting 51 GRAAMs of RMI-related projects. The \textsc{ArCode} learning curve indicates that:

\begin{figure}
    \centering
    \begin{subfigure}[b]{0.48\textwidth}
        \centering\captionsetup{width=.9\linewidth}%
        \includegraphics[width=\textwidth]{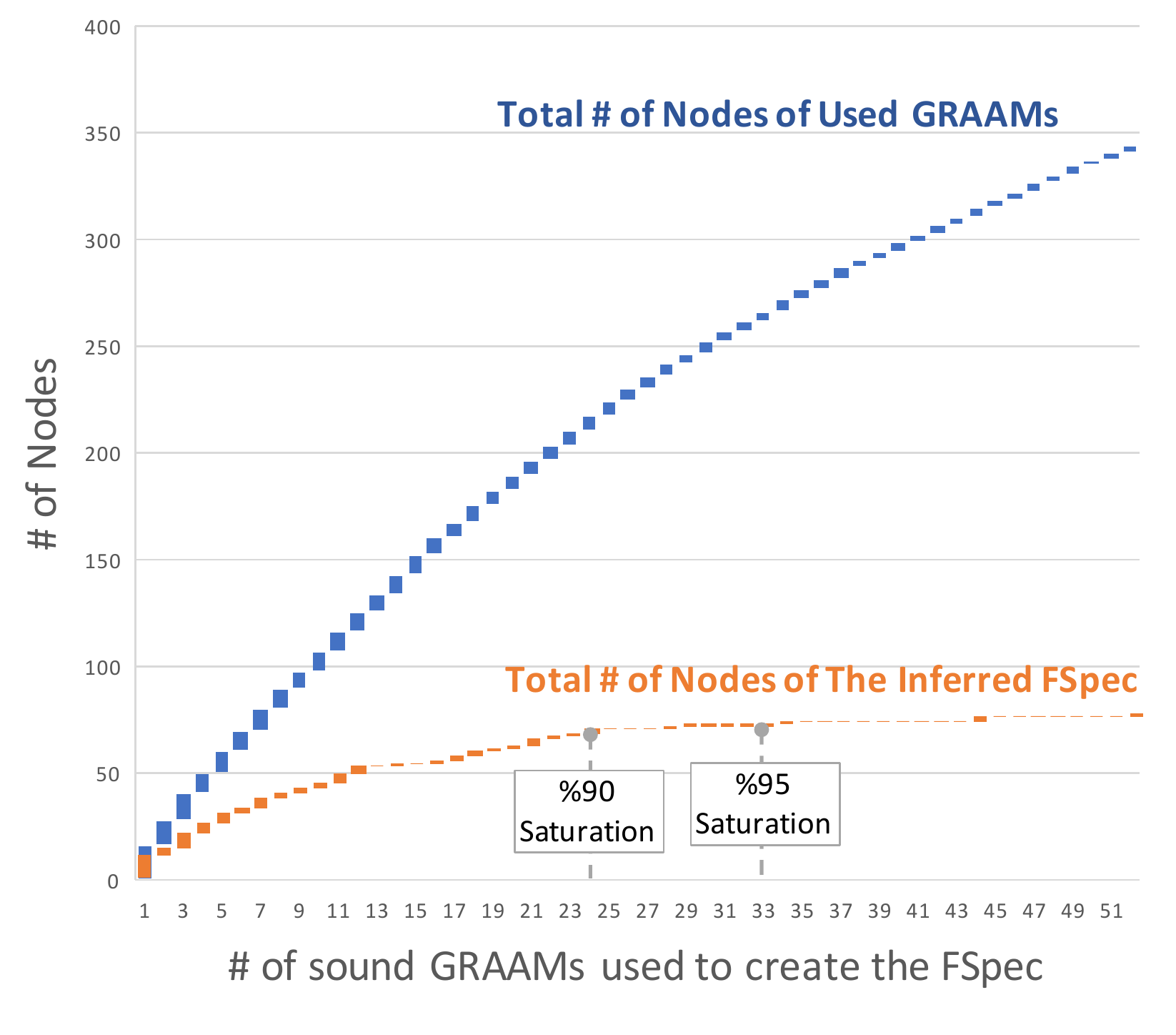}
        \centering
        \caption{ArCode learning curve while visiting 51 JAAS-based sound GRAAMs}
        \label{fig:model_saturation_JAAS}
    \end{subfigure}%
    \begin{subfigure}[b]{0.48\textwidth}
        \centering\captionsetup{width=.9\linewidth}%
        \includegraphics[width=\textwidth]{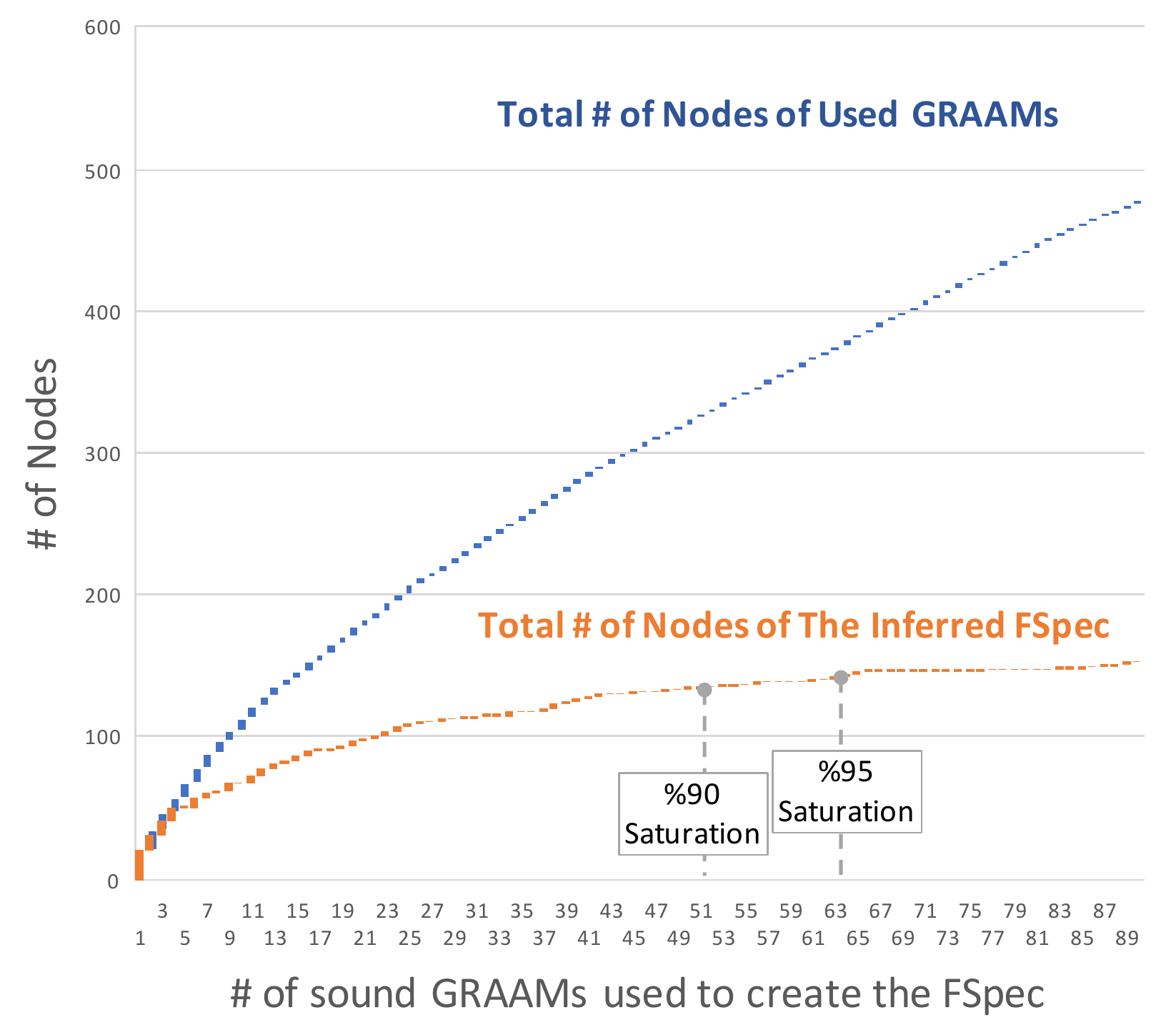}
        \centering
        \caption{ArCode learning curve while visiting 89 RMI-based sound GRAAMs}
        \label{fig:model_saturation_RMI}
    \end{subfigure}%
    \caption{Model saturation while creating FSpecs of JAAS and RMI frameworks}
    \label{fig:model_saturation}
    \vspace{-5pt}
\end{figure}

\begin{tcolorbox}[size=title, opacityfill=0.15]
\textbf{Finding 1:}
Two case studies of RMI and JAAS indicate that there are \textit{limited correct ways} that developers incorporated  these frameworks in their program;  \textsc{ArCode}'s learning technique was able to enumerate and learn all these  possible sound usages of  the frameworks.
\end{tcolorbox}

\subsection{API Recommendation to Implement Tactics/Patterns} \label{sec:eval_API_Recommendation}
To evaluate the quality of recommendations, we randomly select 80\% of the projects of each framework in the code repository as training and the remaining 20\% as testing data set. We use the same train and test data sets for \textsc{ArCode}, GrouMiner, and MAPO approaches to make a fair comparison between their performance.  

To generate labeled test cases for this experiment, we first create GRAAMs of each program in the test data set and then, we remove the last node of each GRAAM. Since a recommendation system is expected to recommend the removed API, that API would be used as the label of that test case. Next, we ask the recommendation system to return a ranked list of recommendations for each test case. Finally, based on the position of the correct recommendation in the ranked list, we compute the top-K accuracy of the recommendations ($k:1\rightarrow10$). Though, since we expect only one correct answer for each test case, the results for top-K precision and top-K recall would be the same as top-K accuracy in our experiments.

\begin{figure}
    \centering
    \begin{subfigure}[b]{0.48\textwidth}
        \centering
        \includegraphics[width=\textwidth]{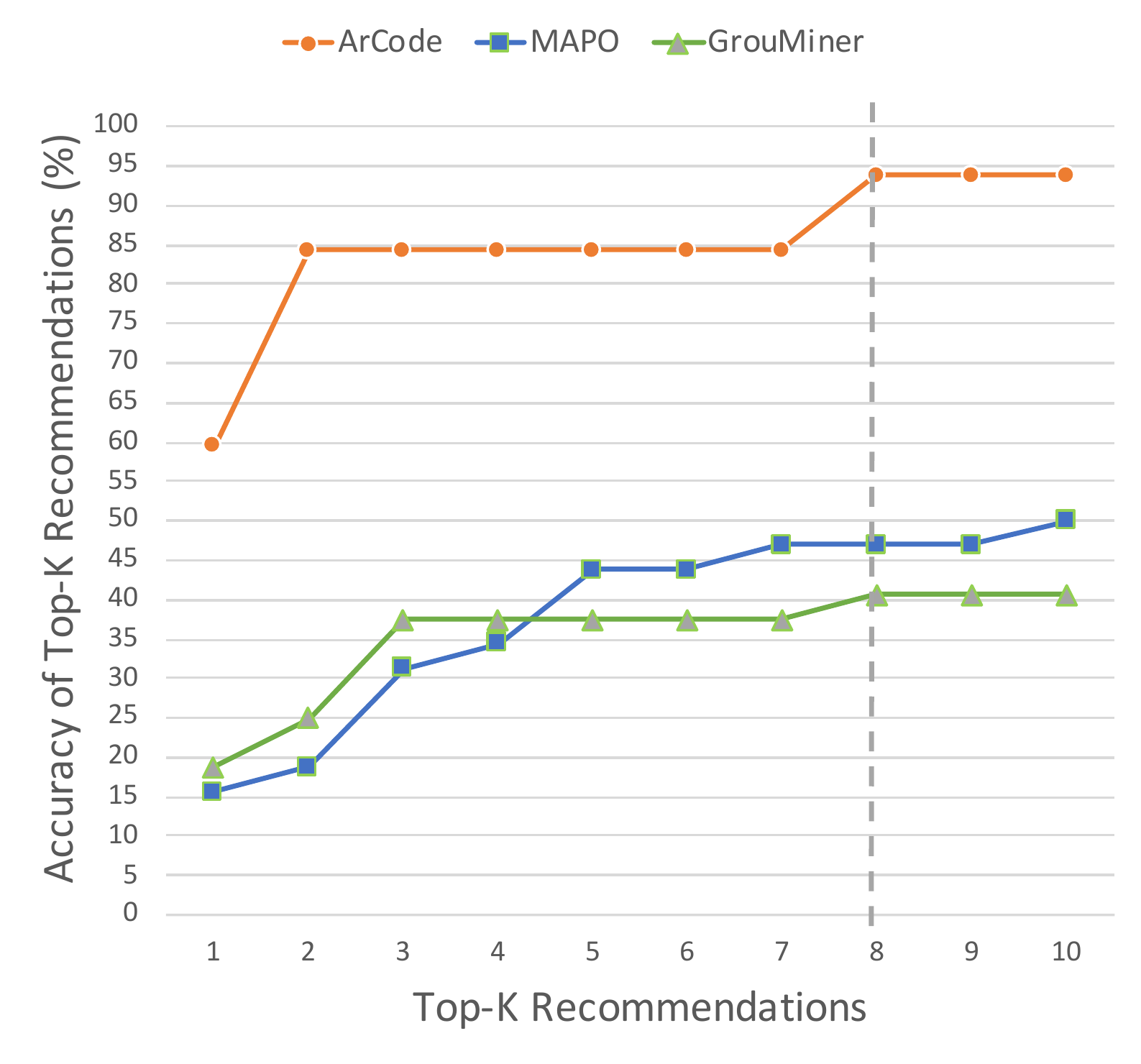}
        \centering
        \caption{JAAS-based projects}
        \label{fig:evaluation_next_API_JAAS}
    \end{subfigure}%
    \begin{subfigure}[b]{0.48\textwidth}
        \centering
        \includegraphics[width=\textwidth]{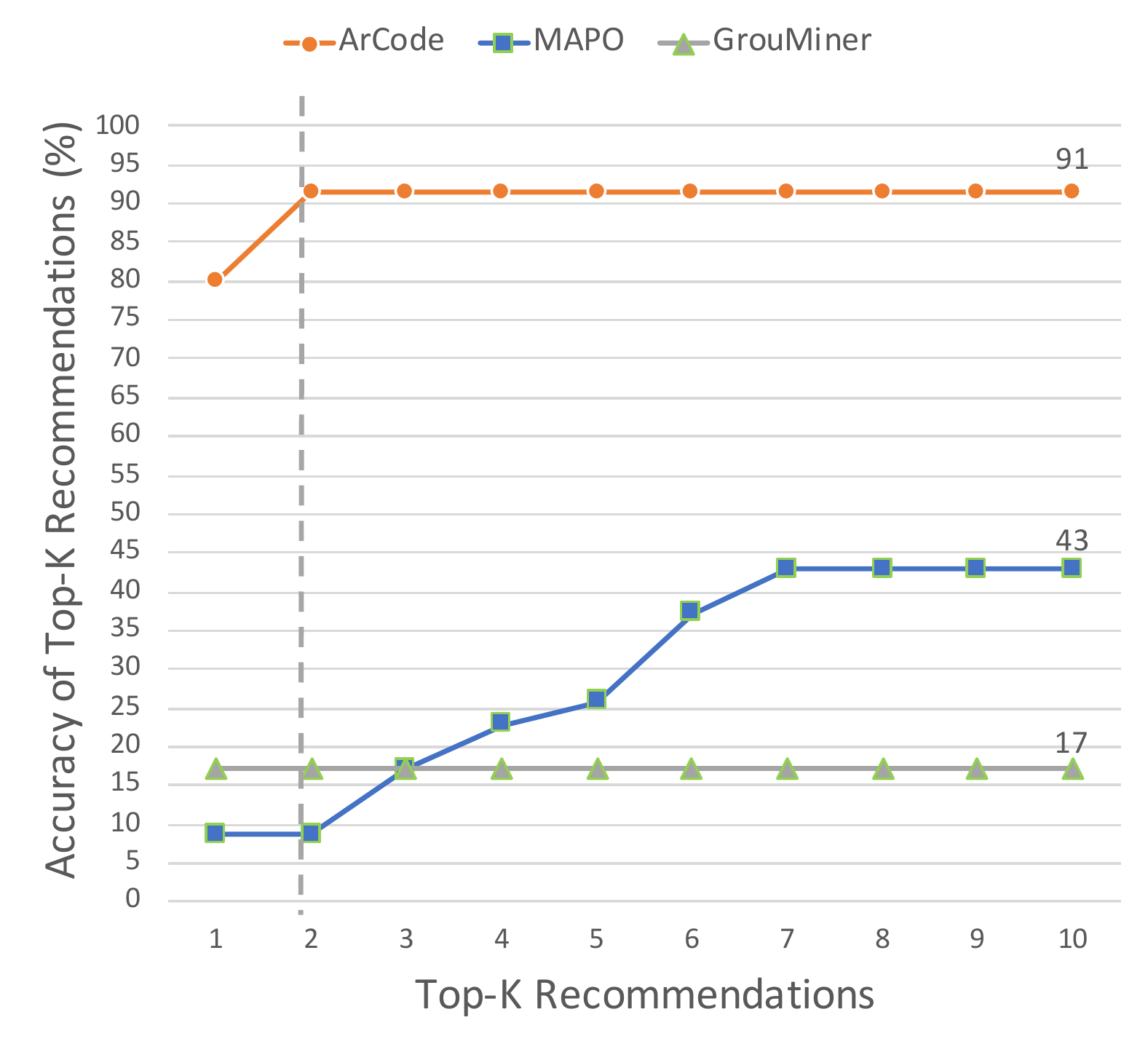}
        \centering
        \caption{RMI-based projects}
        \label{fig:evaluation_next_API_RMI}
    \end{subfigure}%
    \caption{Accuracy of \textsc{ArCode}, GrouMiner, and MAPO for next API recommendations on JAAS and RMI experiments}
    \label{fig:evaluation_next_API}
    \vspace{-5pt}
\end{figure}

We also computed the top-K accuracy of recommendations for MAPO and GrouMiner approaches. \figref{fig:evaluation_next_API} shows the result of this experiment. In the case of JAAS framework (\figref{fig:evaluation_next_API_JAAS}), \textsc{ArCode} achieved a 59\% accuracy while  considering only the top ranked (top-1) API recommendation. However, if we consider top-2 recommendations, the accuracy improves to 84\%. Finally, if we consider top-8 APIs (or beyond), the accuracy of recommendations provided by \textsc{ArCode} tops 94\%. 
Compared to \textsc{ArCode}, the accuracy of MAPO and GrouMiner reaches 50\% and 41\% for JAAS-based programs. For RMI-based programs, ArCode provides 91\% accuracy for top-2 recommendations and beyond. The highest accuracy for MAPO and GrouMiner is 43\% and 17\%. Based on our observation, the diversity of API usages in JAAS-based programs is fewer compared to those of RMI-based programs. As a result, all approaches have their best performance in JAAS-based repository. Nevertheless, \textsc{ArCode} still outperforms MAPO and GrouMiner in both JAAS- and RMI-based tests.

These results bring us to the following observation:

\begin{tcolorbox}[size=title, opacityfill=0.15]
\textbf{Finding 2:}
ArCode's next API recommendation outperforms the prior work significantly (40\% and more). In both case studies top-2 recommendations were reliable (85\% in JAAS and 95\% in RMI), however ArCode top-1 recommendations in RMI were more reliable than JAAS.
\end{tcolorbox}

\subsection{API Misuse: Detecting a Missed API}
One of common API missuses is missing a critical API call while implementing a tactic. Authenticating users with \code{LoginContext.login()} without first calling \code{HttpSession.invalidate()}~\cite{ICSA2017} or checking the role of a user before granting access are just some of many examples.

To examine the accuracy of our approach in identifying such cases of API misuse in programs, similar to the API recommendation experiment (\secref{sec:eval_API_Recommendation}), we generate labeled test cases for this experiment. 
To do so, we randomly remove an API call from each program in an iterative manner. We were able to generate 77 test cases for JAAS-based and 80 test cases for RMI-based experiments. Please note that these test cases are generated only from test projects. Then, we ask the system to identify the missed API and recommend a fix for it. 



\begin{figure}
    \centering
    \begin{subfigure}[b]{0.48\textwidth}
        \centering
        \includegraphics[width=\textwidth]{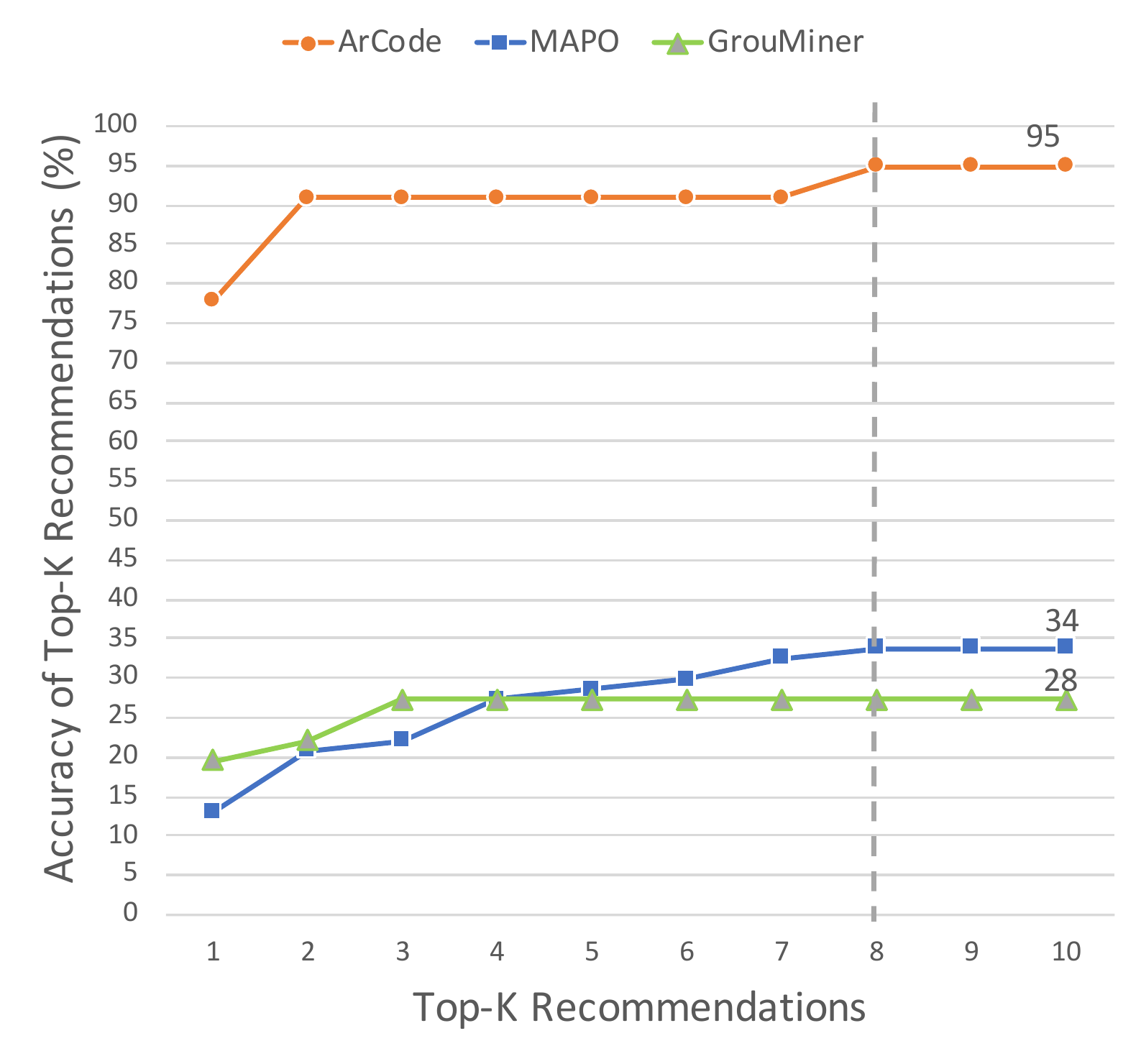}
        \centering
        \caption{JAAS-based projects}
        \label{fig:evaluation_missed_API_JAAS}
    \end{subfigure}%
    \begin{subfigure}[b]{0.48\textwidth}
        \centering
        \includegraphics[width=\textwidth]{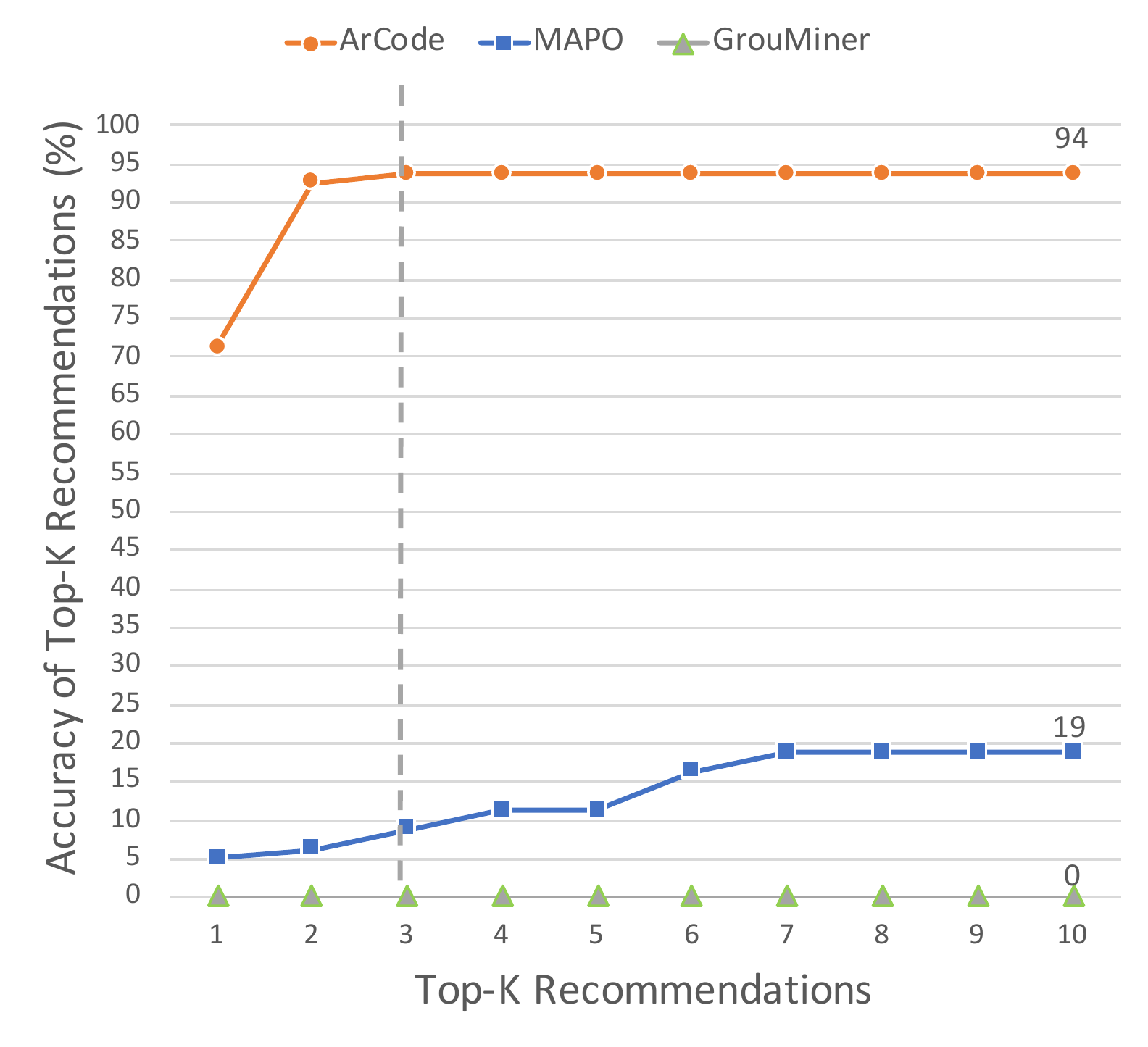}
        \centering
        \caption{RMI-based projects}
        \label{fig:evaluation_missed_API_RMI}
    \end{subfigure}%
    \caption{Accuracy of \textsc{ArCode}, GrouMiner, and MAPO for missed API recommendations on JAAS and RMI experiments}
    \label{fig:evaluation_missed_API}
    \vspace{-5pt}
\end{figure}

\figref{fig:evaluation_missed_API} depicts the result of this experiment. In the case of the JAAS framework (\figref{fig:evaluation_missed_API_JAAS}), \textsc{ArCode} achieved a 78\% accuracy considering only the top ranked (top-1) API recommendation. Top-2 recommendation shows 91\% accuracy and finally, the accuracy of recommendations provided by \textsc{ArCode} tops 95\% for top-8 recommendations and beyond. Comparing to \textsc{ArCode}, the accuracy of MAPO and GrouMiner reaches 34\% and 28\% respectively. This test scenario is more complicated compared to the previous test case (next API recommendation) because, in addition to the ancestors, the descendants of the provided recommendation and the correct answer should match as well. Therefore, the accuracy of pattern-based approaches decreases in this usecase. 
%

Based on the results of this experiment we found:

\begin{tcolorbox}[size=title, opacityfill=0.15]
\textbf{Finding 3:}
ArCode can identify a missed API in implementation of tactics/patterns and provide a recommendation to fix it. ArCade's top-2 recommendation accuracy in JAAS and RMI case studies are above 90\%. Furthermore, ArCode outperforms prior work with 60\% and more, making it a more reliable approach for API recommendations to implement tactics and patterns using frameworks. 
\end{tcolorbox}

\subsection{API Misuse: Wrong API Usage}
Another type of misusing APIs in a program includes calling APIs in a wrong order in that program. Although such a program might not show a compile time error, the expected tactic is not implemented correctly.
In this experiment, we create test cases that include an incorrect API usage in each project (e.g. incorrect order of APIs). Then, we ask the system to first, identify this API misuse and second, recommend a fix to make that API usage correct. 
To create such test cases, we go over all APIs in each program iteratively, selecting two random APIs and swapping them to generate incorrect usages of APIs. Following this approach, we were able to generate 432 test cases for JAAS and 246 test cases for RMI framework. Then, we ask the system to identify this API misuse and recommend a fix for it. Like the two other test scenarios, the system returns a ranked recommendation and we compute accuracy for top-k ($k: 1\rightarrow10$) recommendations. 

\begin{figure}
    \centering
    \begin{subfigure}[b]{0.48\textwidth}
        \centering
        \includegraphics[width=\textwidth]{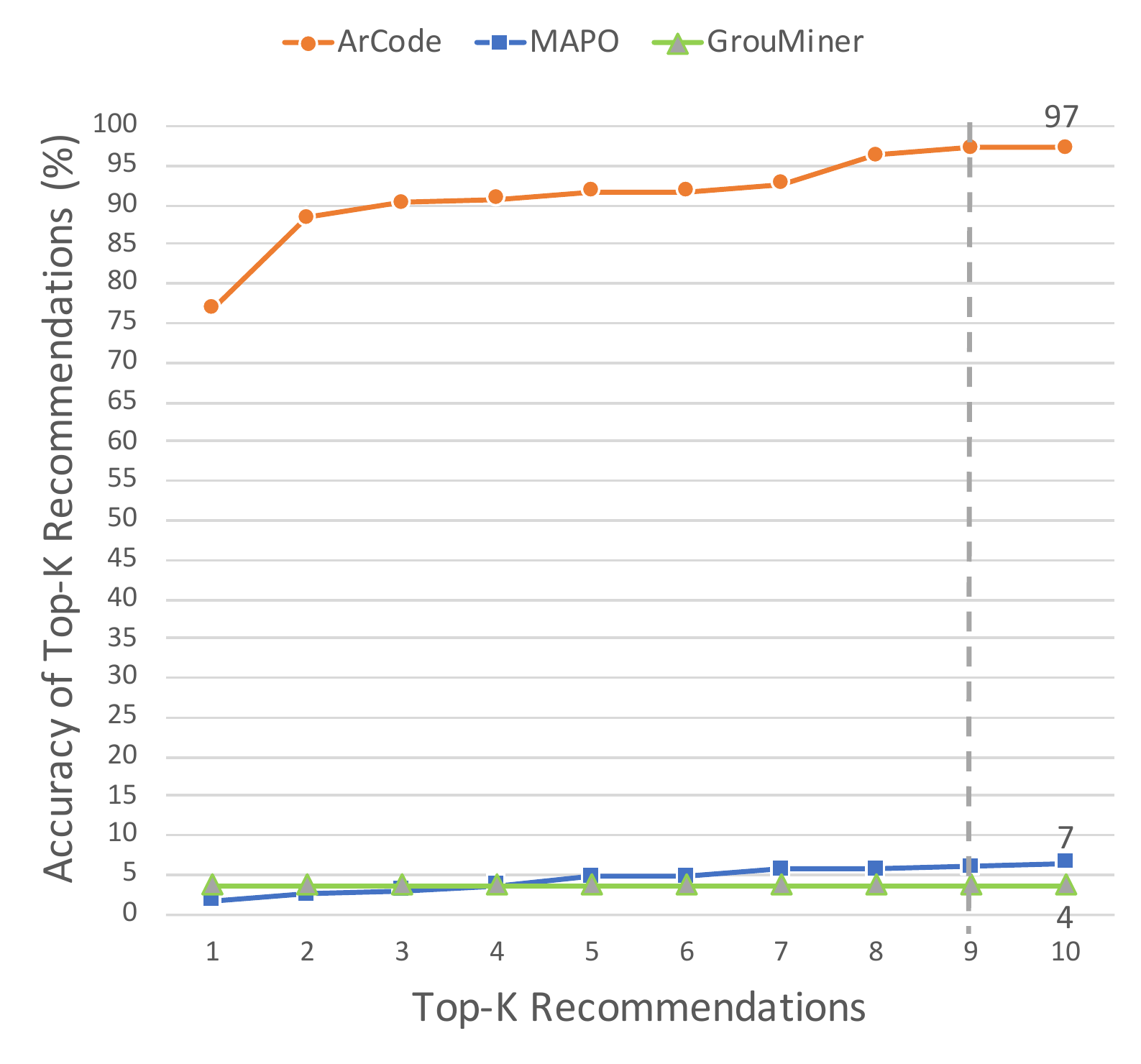}
        \centering
        \caption{JAAS-based projects}
        \label{fig:evaluation_swapped_API_JAAS}
    \end{subfigure}%
    \begin{subfigure}[b]{0.48\textwidth}
        \centering
        \includegraphics[width=\textwidth]{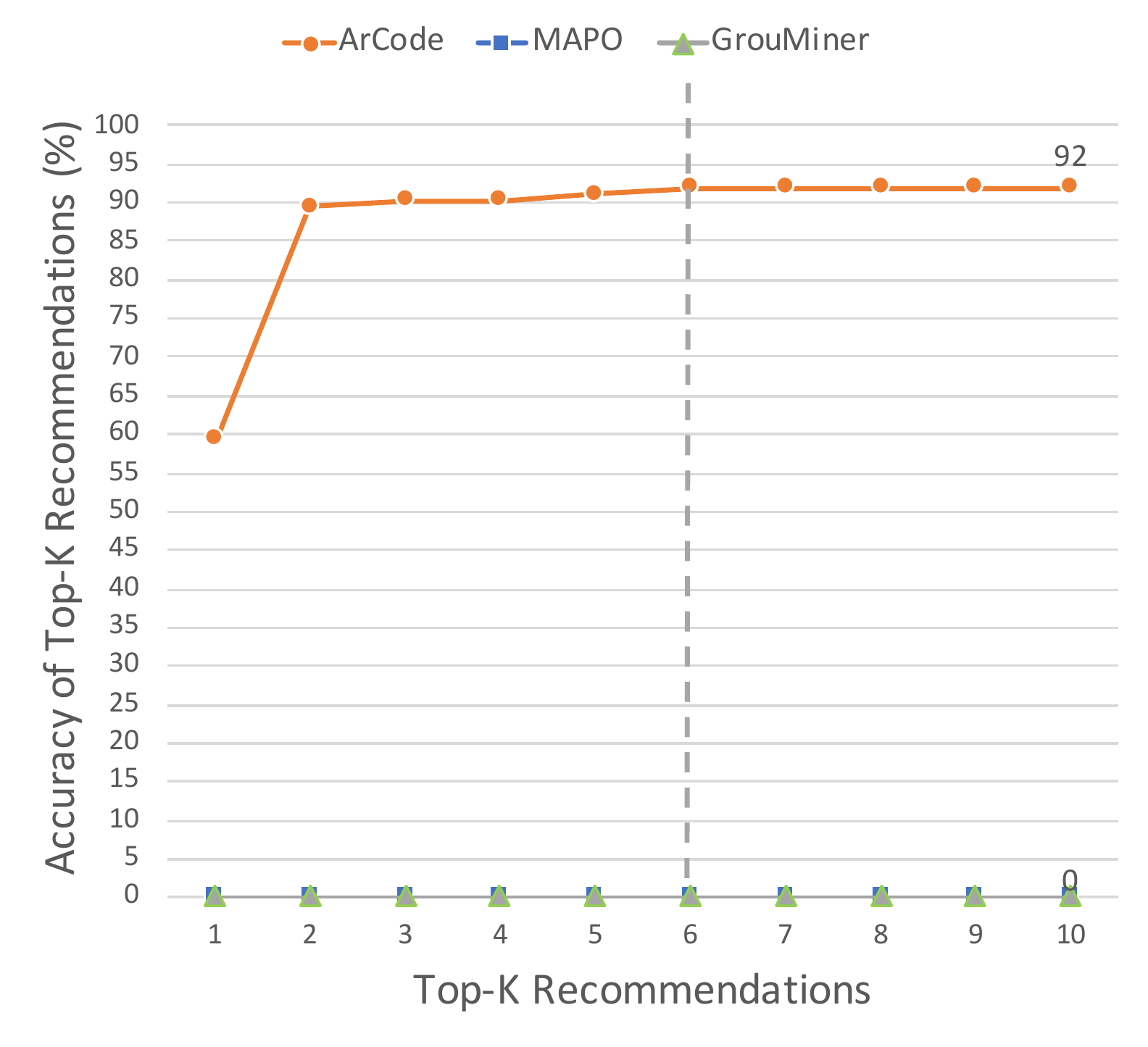}
        \centering
        \caption{RMI-based projects}
        \label{fig:evaluation_swapped_API_RMI}
    \end{subfigure}%
    \caption{Accuracy of \textsc{ArCode}, GrouMiner, and MAPO for fix recommendations on JAAS and RMI experiments}
    \label{fig:evaluation_swapped_API}
    \vspace{-5pt}
\end{figure} 

The result of this experiment is demonstrated in \figref{fig:evaluation_swapped_API}. In the case of JAAS framework, \textsc{ArCode} shows 77\% accuracy for top-1 recommendations. Also, if we consider 8-top returned recommendations (or beyond), the accuracy of \textsc{ArCode} tops 97\%. The results show that MAPO and GrouMiner reach 7\% and 4\% accuracy respectively. This test scenario is the most complicated case compared to the previous experiments. Since we are swapping two APIs, ancestors and descendants of both the first and second APIs should be matched against those that are recommended. Thus, we observe accuracy degrading for MAPO and GrouMiner in such experiments. Based on the results of this experiment, we observed that:

\begin{tcolorbox}[size=title, opacityfill=0.15]
\textbf{Finding 4:}
ArCode accurately detects wrong API usages and generates fix recommendation for them. Wrong API usages are more complicated that finding a missed API call in a tactic/pattern. In these complex misuses, ArCode with a large difference (60\% to 90\%) outperformed two of the prior work (GrouMiner and MAPO). 
\end{tcolorbox}


%% file: ThreatsToValidity.tex
\section{Threats to Validity}\label{sec:Threats_to_Validity}
\textsc{ArCode} aims to filter out incorrect API usages from a code repository and then, learn correct API usages from the remained programs. While the learnt framework API specification model (FSpec) does not include incorrect API usages, we can not claim that it covers all the possible correct API usages. Another notable point is that \textsc{ArCode} leverages conservative merging rules while creating an FSpec. Although it can guarantee that no incorrect API usage emerges in the final FSpec, it can affect the efficiency of the training phase as well as the size of the final FSpec adversely.  
Moreover, the Inter-framework Dependency (IFD) model introduced in this paper is created based on static analysis over the source code of the framework to find reader-writer roles of API methods inside that framework. However, some modern frameworks use dynamic features of object oriented languages (e.g. reflections) to implement the framework. In these cases, performing dynamic analysis alongside  static analysis would result in capturing more accurate dependencies between APIs. 

Concerning the evaluation of \textsc{ArCode} provided in this paper, we used two popular Java based frameworks, JAAS and RMI. For generality purposes, conducting experiments with more frameworks would be advised. 

%% file: RelatedWork.tex
\section{Related Work}\label{sec:relatedWork}

There have been numerous works on identifying frameworks' API specifications and creating models that enable API recommendation and misuse detection systems. The aim of these systems is to help programmers correctly use API calls and prevent incorrect usage of APIs in their programs~\cite{robillard2013automated, zhong2017empirical}. 
Many researchers define patterns of \textbf{API co-occurrences} in the same scope of a program as an API specification. Works in this category find the frequency of co-appearance of APIs in the same context and use it to create API co-occurrence patterns. While most of the approaches in this category find such patterns within a  method~\cite{li2005pr, zheng2011cross, Moritz2013, monperrus2013detecting, saied2015mining, saied2015could, thung2016api, nguyen2016api, xie2019hirec, niu2017api}, some go beyond and find patterns in a bigger scope (e.g. class or program)~\cite{lamba2015pravaaha, nguyen2019focus}.    
We have found these works useful for simple API recommendation or misuse detection tasks. However, they become impractical when the problem gets more complex as to accurately \textbf{(i)} identify an incorrect API \textit{order} in a code, or \textbf{(ii)} recommend the \textit{next} API based on the order of other APIs in a \textit{sequence}.

To address the aforementioned issues, more sophisticated approaches find \textbf{API sequence} patterns in a code repository. This is an important specification needed for a correct API recommendation or misuse detection. The developed techniques in this category range from finding partial API usage patterns and creating rule-based specifications \cite{liu2006ltrules, acharya2007mining, wasylkowski2007detecting} to mining a complete sequence of API usage patterns~\cite{shoham2008static, zhong2009mapo, hsu2011macs, zhang2012automatic, wang2013mining, asaduzzaman2014cscc, dang2015api, saied2020towards,saied2018towards, fowkes2016parameter}. As an example, MAPO~\cite{zhong2009mapo}, one of the most respected API specification miners, aims to create patterns of API usage sequences and leverage that to find relevant code samples for programmers.
Although sequential API specification miners take API order into account, they are not able to: \textbf{(i)} provide more API-related information in a program (e.g. data dependency) than the order of appearance in the code, and \textbf{(ii)} distinguish between two semantically equivalent sequences (\secref{sec:GRAAM}). 

\textbf{Graph-based API} pattern finders are the most advanced methods developed to cover sequence-based API pattern miners shortcomings and create more detailed API specifications  (e.g. data dependency between APIs). Specifically, these works aim to track the usage of APIs related to a single object~\cite{Nguyen2009, nguyen2010graph, nguyen2012grapacc, ghafari2014towards, mover2018mining} or multiple objects~\cite{asaduzzaman2017recommending,asaduzzaman2017femir, pacheco2019mining, amann2019investigating, gu2019codekernel} in a specific scope in programs. For instance, GrouMiner~\cite{Nguyen2009}, a well-appreciated graph-based API pattern miner, finds relationships between APIs of the same type in a method.
While the mentioned works present more detailed API specifications, they still: \textbf{(i)} do not track dependencies across different scopes (inter-procedural), \textbf{(ii)} can not represent the whole context of an API usage, and \textbf{(iii)} can not identify semantically equivalent, yet in different order, API usages in programs.

\textsc{\textbf{ArCode}} performs an inter-procedural program analysis and infers a context-sensitive graph-based API specification for frameworks. This approach produces isomorphically the same graphs for API usages that have different sequences but are semantically equivalent.

%% file: Conclusion.tex
\section{Conclusion \& Future Work}
\label{sec:conclusion}

Obtaining API specification of a framework can enable the correct implementation of architectural tactics and patterns. This paper introduces \textsc{ArCode}, a novel approach for inferring frameworks' API specification from limited projects. It relies on a Graph-based Framework API Usage model (GRAAM), which is an inter-procedural  context-, and flow-sensitive representation of APIs in a program. Furthermore, \textsc{ArCode} also extracts inter-framework dependencies between APIs from framework source code and uses them to identify incorrect API usages in programs. It uses an inference algorithm to combine all extracted GRAAMs into a framework specification model. In a series of experiments, we demonstrated that it is possible to infer a framework specification model that accurately captures correct API usage to implement tactics and patterns. Moreover, recommendation systems empowered by the created framework specification model are able to provide accurate API recommendations, identify API misuses and provide a fix recommendation. 
Future work includes exploration of additional frameworks with regard to our technique and leveraging dynamic analysis alongside static analysis to extracting inter-framework dependencies.